\documentclass[jcappub,11pt]{article}
\pdfoutput=1 
\usepackage{jcappub} 
\usepackage[T1]{fontenc} 
\usepackage{url}
\usepackage{xcolor}
\usepackage{graphicx}
\usepackage{mathrsfs}
\usepackage[caption=false]{subfig}

\usepackage{tikz,xcolor,hyperref}

\definecolor{lime}{HTML}{A6CE39}
\DeclareRobustCommand{\orcidicon}{%
	\begin{tikzpicture}
	\draw[lime, fill=lime] (0,0) 
	circle [radius=0.16] 
	node[white] {{\fontfamily{qag}\selectfont \tiny ID}};
	\draw[white, fill=white] (-0.0625,0.095) 
	circle [radius=0.007];
	\end{tikzpicture}
	\hspace{-5mm}
}

\foreach \x in {A, ..., Z}{%
	\expandafter\xdef\csname orcid\x\endcsname{\noexpand\href{https://orcid.org/\csname orcidauthor\x\endcsname}{\noexpand\orcidicon}}
}

\title{\boldmath Probing massive neutrinos and modified gravity with redshift-space morphologies and anisotropies of large-scale structure}

\author[a,b]{Wei Liu\orcidA{},}
\author[a,b]{Liang Wu,}
\author[c,d]{Francisco Villaescusa-Navarro,}
\author[e,f,g]{Marco Baldi,}
\author[h,i]{Georgios Valogiannis,}
\author[a,b,1]{and Wenjuan Fang \note{Corresponding author.}}

\affiliation[a]{Department of Astronomy, University of Science and Technology of China, Hefei, Anhui, 230026, P.R.China}
\affiliation[b]{School of Astronomy and Space Sciences, University of Science and Technology of China, Hefei, Anhui, 230026, P.R.China}
\affiliation[c]{Center for Computational Astrophysics, Flatiron Institute, 162 5th Avenue, New York, NY 10010, USA}
\affiliation[d]{Department of Astrophysical Sciences, Princeton University, Peyton Hall, Princeton, NJ 08544-0010, USA}
\affiliation[e]{Dipartimento di Fisica e Astronomia, Alma Mater Studiorum - University of Bologna, Via Piero Gobetti 93/2, 40129 Bologna BO, Italy}
\affiliation[f]{INAF - Osservatorio Astronomico di Bologna, Via Piero Gobetti 93/3, 40129 Bologna BO, Italy}
\affiliation[g]{INFN - Istituto Nazionale di Fisica Nucleare, Sezione di Bologna, Viale Berti Pichat 6/2, 40127 Bologna BO, Italy}
\affiliation[h]{Department of Astronomy \& Astrophysics, University of Chicago, Chicago, IL, 60637, USA}
\affiliation[i]{Kavli Institute for Cosmological Physics, Chicago, IL, 60637, USA}

\emailAdd{lw980228@mail.ustc.edu.cn}
\emailAdd{wul@mail.ustc.edu.cn}
\emailAdd{villaescusa.francisco@gmail.com}
\emailAdd{marco.baldi5@unibo.it}
\emailAdd{gvalogiannis@uchicago.edu}
\emailAdd{wjfang@ustc.edu.cn}

\abstract{Strong degeneracy exists between some modified gravity (MG) models and massive neutrinos because the enhanced structure growth produced by modified gravity can be suppressed due to the free-streaming massive neutrinos. Previous works showed this degeneracy can be broken with non-Gaussian or velocity information. Therefore in this work, we focus on the large-scale structure (LSS) in redshift space and investigate for the first time the possibility of using the non-Gaussian information and velocity information captured by the 3D scalar Minkowski functionals (MFs) and the 3D Minkowski tensors (MTs) to break this degeneracy. Based on the Quijote and Quijote-MG simulations, we find the imprints on redshift space LSS left by the Hu-Sawicki $f(R)$ gravity can be discriminated from those left by massive neutrinos with these statistics. With the Fisher information formalism, we first show how the MTs extract information with their perpendicular and parallel elements for both low- and high-density regions; then we compare constraints from the power spectrum monopole and MFs in real space with those in redshift space, and investigate how the constraining power is further improved with anisotropies captured by the quadrupole and hexadecapole of the power spectrum and the MTs; finally, we combine the power spectrum multipoles with MFs plus MTs and find the constraints from the power spectrum multipoles on $\Omega_{\mathrm{m}}, h,  \sigma_8$, $M_\nu$, and $f_{R_0}$ can be improved, because they are complemented with non-Gaussian information, by a factor of 3.4, 3.0, 3.3, 3.3, and 1.9 on small scales ($k_{\rm{max}}=0.5~h\rm{Mpc}^{-1},\ R_G=5~h^{-1}\rm{Mpc}$), and 2.8, 2.2, 3.4, 3.4, and 1.5 on larger scales ($k_{\rm{max}}=0.25~h\rm{Mpc}^{-1},\ R_G=10~h^{-1}\rm{Mpc}$).}

\begin{document}
\maketitle
\flushbottom

\section{Introduction}
The concordance $\Lambda$ cold dark matter ($\Lambda$CDM) model has been used to explain the observed Universe during the past $\sim30$ years, which uses General Relativity (GR) to describe gravitational interaction \cite{2003ApJS..148..175S,2013ApJS..208...19H,2020A&A...641A...6P,2021PhRvD.103h3533A,2022PhRvD.105b3520A}. However, modifications to GR have been considered to explain the late-time acceleration (see \cite{2024arXiv241112026I} for a current modified gravity constraint from DESI \cite{2016arXiv161100036D}), such as the Hu-Sawicki f(R) gravity model \cite{2007PhRvD..76f4004H}. A key feature of f(R) gravity is that the growth of cosmic structure is enhanced in a scale-dependent way, which can be utilized to probe departures from general relativity. However, the structure growth can be suppressed by free-streaming neutrinos \cite{2006PhR...429..307L,2011ARNPS..61...69W,10.1111/j.1365-2966.2011.19488.x},
whose mass is observed to be non-zero \cite{PhysRevLett.81.1562,PhysRevLett.89.011301,PhysRevLett.94.081801,PhysRevLett.101.131802}. A strong degeneracy is reported between modified gravity and massive neutrinos in \cite{PhysRevD.88.103523,PhysRevLett.110.121302}, which is found to exist in three traditional statistics of LSS \cite{10.1093/mnras/stu259}: the power spectrum, the halo mass function, and the halo bias. The authors of \cite{2019MNRAS.486.3927H} took a closer look at the halo mass function and concluded this degeneracy is a limiting factor for the current telescopes. 

Alternative statistics can be used to explore unique imprints left on LSS by each model and discriminate the f(R) gravity and massive neutrinos. Based on the DUSTGRAIN-pathfinder N-body simulations, which have implemented simultaneously the effects of f(R) gravity and massive neutrinos, weak lensing tomographic information captured by the power spectrum and one-point statistics at multiple redshifts was found to be very promising in distinguishing some of these degenerate non-standard models from $\Lambda$CDM \cite{10.1093/mnras/sty2465}. This work is followed by a series of papers: \cite{2018A&A...619A..38P} investigated various higher-order statistics of the weak-lensing signal to break degeneracies between massive neutrinos and MG; \cite{10.1093/mnras/stab1112} found that the enhancement in the void density profiles can be almost completely overridden by massive neutrinos, but the void size function at high redshifts and for large voids may be effective in disentangling the degeneracy; \cite{Lee_2022} detected the difference between models with degenerate traditional statistics of LSS using the turnaround radii of dark matter halos; the intrinsic shape alignments of massive halos were found to differ substantially for these models \cite{Lee_2023}; the drifting coefficient of the field cluster mass function can discriminate with high statistical significance the degenerate models \cite{2020ApJ...904...93R}.

On the other hand, the information on structure growth embedded in redshift-space distortions (RSD) \cite{1998ASSL..231..185H}, if fully exploited, could help break the degeneracy between MG parameters and neutrino masses. Based on N-body simulations with either f(R) gravity or GR and with massive neutrinos, \cite{2019A&A...629A..46H} studied both the velocity divergence power spectrum and velocity dispersion inside clusters and found the velocity information is powerful in distinguishing massive neutrinos from modified gravity. This work is closely followed by \cite{10.1093/mnras/stz1850}, where the RSD information was extracted with the multipole moments of the two-point correlation function, and a Bayesian analysis based on mock halo catalogs showed this anisotropic information is effective in disentangling degeneracies. Almost at the same time, \cite{Wright_2019} modeled the effects of both MG and massive neutrinos on real- and redshift-space power spectra with standard perturbation theory, they found that the velocity information contained in the quadrupole of the redshift-space power spectrum is sufficient to distinguish GR with light neutrinos and f(R) with heavy neutrinos.

In this work, for the first time, we try to quantitatively answer how the non-Gaussian information captured by the morphological properties of LSS, which are fully characterized by 4 Minkowski functionals (MFs) \cite{1994A&A...288..697M}, and the anisotropic information in redshift space extracted by the generalization of Minkowski functionals to Minkowski tensors (MTs)\cite{schroder2011minkowski,2013NJPh...15h3028S} help break the f(R) gravity and massive neutrinos degeneracy and tighten constraints on the f(R) parameter, neutrino masses and other cosmological parameters.

According to Hadwidger's theorem \cite{Hadwiger_1957}, for a pattern in n-dimensional space, its morphological properties (defined as those satisfying motional-invariance and additivity) can be fully described by (n+1) Minkowski functionals \footnote{The more rigorous and mathematical description of Hadwiger's theorem is given in \cite{1994A&A...288..697M} as: any additive, motion invariant and conditionally continuous functional $\mathscr{F}$ on a body $A$ in $d$ dimension is a linear combination of the $d+1$ Minkowski functional $\mathscr{F}(A)=\sum_{\i=0}^d c_i V_i(A)$, with real coefficients $c_i$ independent of $A$.}. In 3D, the 4 MFs are, respectively, the pattern's volume, surface area, integrated mean curvature, and Euler characteristic (or genus). The MFs can principally probe all orders of statistics \cite{1994A&A...288..697M,10.1046/j.1365-8711.1999.02912.x}. Since their introduction into cosmology by \cite{1994A&A...288..697M} in the 1990s, they have been applied to examine the Gaussianity of primordial perturbations \citep[see e.g.,][]{10.1111/j.1365-2966.2012.21103.x,2013MNRAS.435..531C,2006ApJ...653...11H,2013PhRvD..88d1302F}, to test theories of gravity \citep[see e.g.,][]{2017PhRvL.118r1301F,Shirasaki+17}, to probe the reionization epoch \citep[e.g.][]{Gleser+06,Chen+19}, and to investigate neutrino mass in addition to several other interesting topics \cite{2001ApJ...551L...5S,2001A&A...379..412B,2012PhRvD..85j3513K,2005ApJ...633....1P}. In particular, the strong constraining power of MFs on modified gravity was first reported and interpreted in \cite{2017PhRvL.118r1301F}; then \cite{2023arXiv230504520J} quantified the information about f(R) gravity encoded in the MFs by performing a Fisher forecast based on the dark matter and dark matter halo field using N-body simulation; \cite{2022JCAP...07..045L} showed the Minkowski functionals of LSS can reveal the imprints of massive neutrinos on LSS, provide important complementary information to two-point statistics, and significantly improve constraints on $M_{\nu}$; \cite{2023JCAP...09..037L} took a step forward and applied the statistics to mock galaxies in redshift space, strong constraining power was reported on $M_{\nu}$ together with other cosmological parameters from the MFs of the redshift-space galaxy distribution for the first time.

The effect of redshift space distortion on the MFs was first studied in \cite{1996ApJ...457...13M}, where they found that these statistics in redshift space have the same shape as in real space, the redshift space distortion only affects amplitudes of these statistics in the Gaussian limit. The non-Gaussian RSD effect was studied in \cite{2013MNRAS.435..531C}, where new prospects were also explored to utilize the anisotropies induced by RSD for the extraction of velocity information. This idea was closely followed in a series of works \cite{2018ApJ...863..200A,2019ApJ...887..128A,2022arXiv220810164A}, where the impact of RSD on the tensor Minkowski functionals was investigated in details, and it was shown the Minkowski tensors are sensitive to global anisotropic signals present within a field, which can be used to place constraints on the redshift-space distortion parameter $\beta=f/b$ ($f$ is the linear growth rate and $b$ is the linear bias). A new statistic with a very close relation to Minkowski functionals was derived in \cite{2023arXiv230803086J} for stronger constraining power on $\beta$. 

For the first time, in this work, we examine whether the distinct imprints on the morphological properties of LSS left by f(R) gravity and massive neutrinos can be detected by the MFs at the same time. Then we utilize the MTs \footnote{In this paper, ``MTs'' refer to the tensor Minkowski functionals only, not including the scalar Minkowski functionals, when we are introducing their definitions, discussing their difference with the scalar MFs, and focusing the extra information embedded in them, etc. But in other cases, ``MTs'' refer to the combination of scalar and tensor Minkowski functionals, since the scalar MFs are just a special kind of MTs.  We believe the reader can understand its meaning according to the context and ambiguities are not introduced.} to further extract the anisotropies introduced by RSD. Distinct signatures of f(R) gravity and massive neutrinos on these statistics are obtained, and then used to place constraints on the present-day value of the background scalar field $f_{R_0}$, the neutrino mass sum $M_{\nu}$, and other cosmological parameters.  We also quantify the information content on these parameters from the power spectrum in real- and redshift-spaces, and compare their differences, as well as examine how its quadrupole and hexadecapole help break the degeneracy and improve parameter constraints. The monopole, quadrupole, and hexadecapole of the power spectrum are also considered as a benchmark to quantify the constraining power of different statistics and the efficiency of the 3D MTs in capturing velocity information. The combination of these statistics is shown to be a promising probe of the f(R) parameter, neutrino mass, and other cosmological parameters.

This paper is organized as follows. In Section \ref{models}, we present the Quijote and Quijote-MG simulation suite. We then describe the statistics used in this work in Section~\ref{statistics}. The Fisher information matrix formalism used to calculate parameter constraints is given in Section~\ref{sec:fisher}, and the forecasted results are presented in Section \ref{sec:result}. Section~\ref{sec:discuss} discusses the results obtained and makes a comparison with previous works.  Finally, we conclude in Section \ref{conclusions}. In the Appendices~\ref{sec:Gaussian_test}, \ref{sec:conver}, \ref{sec:test_deri}, \ref{sec:k-cut}, we discuss some subtleties of our forecast, and present a more conservative result in Appendix~\ref{sec:conservative_result}.

\section{The Quijote and Quijote-MG simulations}
\label{models}
Our analysis is based on the Quijote \cite{2020ApJS..250....2V} and Quijote-MG \cite{2024QuijoteMG} (modified gravity simulations for the Hu \& Sawicki f(R) model \cite{2007PhRvD..76f4004H}) sets of simulations, which are run using the TreePM+SPH code GADGET-III \cite{10.1111/j.1365-2966.2005.09655.x} and MG-GADGET\cite{2013MNRAS.436..348P}, respectively. In a cosmological volume of $1\left(h^{-1} \mathrm{Gpc}\right)^{3}$, $512^3$ CDM particles (plus $512^3$ neutrino particles for cosmologies with massive neutrinos) are evolved from redshift $z = 127$ to $z = 0$. For cosmologies with massive neutrinos and/or modified gravity,  and their fiducial counterparts, the initial conditions (ICs) are generated employing the Zel'dovich approximation (ZA); while for all other cosmologies used in this work, the initial conditions are generated using second-order perturbation theory (2LPT) instead. 

The $f(R)$ gravity model is a straightforward extension of General Relativity \cite{DeFelice2010}, which replaces the Ricci scalar, $R$, in the Einstein–Hilbert action, $S$, with an algebraic function of $R$:
\begin{equation}
S=\int d^4 x \sqrt{-g}\left[\frac{R+f(R)}{16 \pi G}+\mathcal{L}_{\mathrm{m}}\right],
\end{equation}
where $g$ is the determinant of the metric, $G$ is the Newtonian gravitational constant, and $\mathcal{L}_{\mathrm{m}}$ is the Lagrangian density for matter fields. The standard GR expression is recovered when $f(R)=0$. In the weak-field and quasi-static limit, the equations of motion employed in the simulations are a modified Poisson equation plus an equation for the scalar field ($f_R \equiv \mathrm{~d} f(R) / \mathrm{d} R$):
\begin{equation}
\begin{aligned}
\nabla^2 \Phi & =\frac{16 \pi G}{3} \delta \rho_m-\frac{1}{6} \delta R, \\
\nabla^2 f_R & =\frac{1}{3}(\delta R-8 \pi G \delta \rho_m),
\end{aligned}
\end{equation}
where $\delta \rho_m$ and $\delta R$ are the matter density perturbations and the perturbation of the Ricci scalar. Therefore, particles in the f(R) simulations experience an additional fifth force sourced by the scalaron $f_R$. For the Hu \& Sawicki f(R) model studied in this work,
\begin{equation}
f(R)=-m^2 \frac{c_1\left(R / m^2\right)^n}{c_2\left(R / m^2\right)^n+1},
\end{equation}
with $m^2=H^2_0\Omega_{\rm m}$, where $H_0$ denotes the Hubble constant, $\Omega_{\rm m}$ is the present-day fractional matter density evaluated, $c_1$, $c_2$, and $n$ are free parameters of the model. We choose $n=1$ for simplicity in this work. Then, after restricting the model to obtain a background evolution close to that of a $\Lambda$CDM model, the model can be fully specified by the background value of the present scalar degree of freedom $f_{R_0}$.

We want to emphasize two characteristic features of the Hu \& Sawicki f(R) model which will be explored and exploited in this work. First, the so-called chameleon screening mechanism \cite{PhysRevLett.93.171104} is employed in this model. In dense environments like clusters, the chameleon mechanism operates efficiently, suppressing gravity modification. In contrast, in underdense regions such as voids, the chameleon mechanism fails to function, leading to significant gravity modification \cite{2012MNRAS.421.3481L,2011PhRvL.107g1303Z}. We measure the MTs as a function of the density contrast in this work, therefore, the MTs can serve as a natural probe of the environmental dependence of the chameleon screening mechanism, and thus also a natural probe of modified theories of gravity \cite{2017PhRvL.118r1301F}. Second, the f(R) gravity modifies the growth rate of structure \cite{Jennings:2012pt}. Since the statistics in redshift space can be used to measure the growth rate of structure, we will focus on the large-scale structure in redshift space in this work.

The fiducial model has the cosmological parameter values set to be in good agreement with the Planck constraints \cite{2020A&A...641A...6P}: the matter density parameter $\Omega_{\mathrm{m}}=0.3175$, the baryon density parameter $\Omega_{\mathrm{b}}=0.049$, the dimensionless Hubble constant $h=0.6711$, the spectral index $n_{s}=0.9624$, the root-mean-square amplitude of the linear matter fluctuations at $8h^{-1} \rm{Mpc}$ $\sigma_{8}=0.834$, the sum of neutrino masses $M_{\nu}=0.0$ eV, the present background value of scalar degree of freedom for Hu \& Sawicki f(R) model $f_{R_0}=0$, and the dark energy state parameter $w=-1$. For the fiducial model, 15000 realizations are run for the accurate estimate of covariance matrices, while for the fiducial model with Zel'dovich ICs, the four models with modified gravity, the three models with massive neutrinos, and the models where only one of the parameters $\Omega_m,\Omega_b,h,n_s,\sigma_8$ varies at a time, 500 realizations are run to precisely estimate the derivatives w.r.t. these parameters. Specifications of the simulations used in this work can be found in Table \ref{tab:s}, only the snapshots at $z=0$ are analyzed.

\begin{center}
	\small
	\begin{table}[tbp]
	\begin{tabular}{|ccccccccc|}
		\hline Name & $f_{R_0}$ & $M_{\nu}$ & $\Omega_{m}$ & $\Omega_{b}$ & $h$ & $n_{s}$ & $\sigma_{8}$ & ICs \\
		\hline Fiducial & 0.0 & 0.0 & 0.3175 & 0.049 & 0.6711 & 0.9624 & 0.834 & 2LPT\\
		Fiducial ZA & 0.0 & 0.0 & 0.3175 & 0.049 & 0.6711 & 0.9624 & 0.834 & Zel'dovich\\
		$f_{R}^{+}$ & \underline{$-5\times10^{-7}$} & 0.0 & 0.3175 & 0.049 & 0.6711 & 0.9624 & 0.834 & Zel'dovich\\
		$f_{R}^{++}$ & \underline{$-5\times10^{-6}$} & 0.0 & 0.3175 & 0.049 & 0.6711 & 0.9624 & 0.834 & Zel'dovich\\
		$f_{R}^{+++}$ & \underline{$-5\times10^{-5}$} & 0.0 & 0.3175 & 0.049 & 0.6711 & 0.9624 & 0.834 & Zel'dovich\\
		$f_{R}^{++++}$ & \underline{$-5\times10^{-4}$} & 0.0 & 0.3175 & 0.049 & 0.6711 & 0.9624 & 0.834 & Zel'dovich\\
		$M_{\nu}^{+}$ & 0.0 & \underline{0.1} & 0.3175 & 0.049 & 0.6711 & 0.9624 & 0.834 & Zel'dovich\\
		$M_{\nu}^{++}$ & 0.0 & \underline{0.2} & 0.3175 & 0.049 & 0.6711 & 0.9624 & 0.834 & Zel'dovich\\
		$M_{\nu}^{+++}$ & 0.0 & \underline{0.4} & 0.3175 & 0.049 & 0.6711 & 0.9624 & 0.834 & Zel'dovich\\
		$\Omega_{m}^{+}$ & 0.0 & 0.0 & \underline{0.3275} & 0.049 & 0.6711 & 0.9624 & 0.834 & 2LPT\\ 
		$\Omega_{m}^{-}$ & 0.0 & 0.0 & \underline{0.3075} & 0.049 & 0.6711 & 0.9624 & 0.834 & 2LPT\\ 
		$\Omega_{b}^{++}$ & 0.0 & 0.0 & 0.3175 & \underline{0.051} & 0.6711 & 0.9624 & 0.834 & 2LPT\\ 
		$\Omega_{b}^{--}$ & 0.0 & 0.0 & 0.3175 & \underline{0.047} & 0.6711 & 0.9624 & 0.834 & 2LPT\\ 
		$h^{+}$ & 0.0 & 0.0 & 0.3175 & 0.049 & \underline{0.6911} & 0.9624 & 0.834 & 2LPT\\ 
		$h^{-}$ & 0.0 & 0.0 & 0.3175 & 0.049 & \underline{0.6511} & 0.9624 & 0.834 & 2LPT\\ 
		$n_s^{+}$ & 0.0 & 0.0 & 0.3175 & 0.049 & 0.6711 & \underline{0.9824} & 0.834 & 2LPT\\ 
		$n_s^{-}$ & 0.0 & 0.0 & 0.3175 & 0.049 & 0.6711 & \underline{0.9424} & 0.834 & 2LPT\\ 
		$\sigma_{8}^{+}$ & 0.0 & 0.0 & 0.3175 & 0.049 & 0.6711 & 0.9624 &  \underline{0.849} & 2LPT\\ 
		$\sigma_{8}^{-}$ & 0.0 & 0.0 & 0.3175 & 0.049 & 0.6711 & 0.9624 &  \underline{0.819} & 2LPT\\
		\hline
		
	\end{tabular}
    \caption{\label{tab:s} The subsets of the Quijote simulation suites used in this work. 5000 fiducial simulations of the Quijote suite are used for estimating the covariance matrices, and 500 simulations each for 14 different cosmologies are used for calculating derivatives of observables with respect to cosmological parameters.}
    \end{table}
\end{center}

\section{Statistics in redshift space}
\label{statistics}
For each simulation, redshift-space distortions are applied along the z-axis, 
thus, the position of a particle in redshift space is given by
\begin{equation}
	\boldsymbol{s}=\boldsymbol{r}+u_z \hat{z},
\end{equation}
where $\boldsymbol{r}$ is the position in real space and $u_z = \boldsymbol{v} \cdot \hat{z}/aH(a)$, $\boldsymbol{v}$ is the peculiar velocity and $H(a)$ is the Hubble parameter as a function of the scale factor $a$. The fully non-linear expression for the Fourier transform of the density contrast in redshift space is \cite{1999ApJ...517..531S}
\begin{equation}
\label{eq:rtos}
\hat{\delta}_s(\boldsymbol{k})=\int \frac{\mathrm{d}^3 \boldsymbol{r}}{(2 \pi)^3} \mathrm{e}^{-\mathrm{i} \boldsymbol{k} \cdot \boldsymbol{r}} \mathrm{e}^{\mathrm{i} f k_z v_z(\boldsymbol{r})}\left[\delta^{(r)}(\boldsymbol{r})+f \nabla_z v_z(\boldsymbol{r})\right],
\end{equation}
where $v_z=\boldsymbol{v} \cdot \hat{z}$ and $\delta^{(r)}(\boldsymbol{r})$ is the density contrast in real space. In this work, we will use both the multipoles of the power spectrum and the MFs plus MTs to extract cosmological information embedded in the redshift-space CDM particle field. The redshift-space distortions are important because they provide valuable velocity information that is crucial for constraining modified gravity models. In practice, we can’t directly observe the redshift-space CDM particle field; galaxy surveys, for example, only trace it in a biased way. However, we decided to start with the CDM distribution in redshift space to keep the analysis more controlled initially. The galaxy bias and other observational factors that might complicate the analysis will be explored step by step. This work represents the first step towards halos and galaxies in redshift space. Our next goal is to conduct a more realistic analysis based on the redshift-space dark matter halo distribution, though we note that the Fisher forecast for biased tracers may face challenges in convergence due to the higher noise in the derivative of the statistics \cite{2024arXiv240606067W}.

\subsection{The multipoles of the power spectrum}
\label{sec:pkmtpole}
To see how density and velocity information are captured by the multipoles of the power spectrum, let's take a look at the model proposed by \cite{PhysRevD.70.083007},
\begin{equation}
P^s(k, \mu)=\left(P_{\delta \delta}(k)+2 \mu^2 P_{\delta \theta}(k)+\mu^4 P_{\theta \theta}(k)\right) \times \mathrm{e}^{-\left(k \mu \sigma_v\right)^2}
\end{equation}
where $\mu$ is the cosine of the angle between the wave vector $\boldsymbol{k}$ and the line of sight (LoS), $\theta = \Delta \cdot \boldsymbol{u}$, 
$P_{\delta \delta}(k)$, $P_{\theta \theta}(k)$, and $P_{\theta \delta}(k)$ are the density auto-power spectrum, velocity divergence auto-power spectrum, velocity divergence and density cross-power spectrum, and
 $\sigma_v$ is the 1D linear velocity dispersion given by
 \begin{equation}
\sigma_v^2=\frac{1}{3} \int \frac{P_{\theta \theta}(k)}{k^2} \mathrm{~d}^3 k .
\end{equation}
The 2D power spectrum can be decomposed into multipole moments by
\begin{equation}
P_l^s(k)=\frac{2 l+1}{2} \int_{-1}^1 P(k, \mu) L_l(\mu) \mathrm{d} \mu,
\end{equation}
$L_l(\mu)$ is the Legendre polynomial. We measure the monopole, quadrupole, and hexadecapole of the power spectrum using the routines provided in the Python package Pylians \cite{Pylians}. \footnote{Pylians stands for Python libraries for the analysis of numerical simulations. They are a set of Python libraries, written in Python, cython, and C, whose purpose is to facilitate the analysis of numerical simulations (both N-body and hydrodynamic). See more information about Pylians on https://pylians3.readthedocs.io/en/master/index.html} These multipoles are estimated up to $k_{\rm{max}}=0.5~h\rm{Mpc}^{-1}$, with bins of size $k_f=2\pi/L$ ($L$ is the box size of the simulation, $L=1000h^{-1}\rm{Mpc}$). We choose to only include k-modes up to $k_{\rm{max}}=0.5~h\rm{Mpc}^{-1}$, for these modes the difference is smaller than $1\%$ between the power spectrum of higher resolution simulations ($1024^3$ particles in a box of $1(h^{-1}\rm{Gpc})^3$) and that of simulations ($512^3$ particles in a box of $1(h^{-1}\rm{Gpc})^3$) used in this work.

\subsection{The Minkowski functionals and tensors}
\label{sec:3dmts}

For a body $K$ embedded in Euclidean space $\mathbb{R}^3$ and bounded by a sufficiently smooth surface $\partial K$, its Minkowski functionals are defined as 
\begin{equation}
\label{eq:w0}
	W_{0}(K)=\int_KdV,
\end{equation}
and
\begin{equation}
\label{eq:wnu}
	W_{\nu}(K)=\frac{1}{3}\int_{\partial K}G_{\nu}dA,
\end{equation}
where $\nu=1,\ 2,\ 3$, $G_1=1$, the mean curvature $G_2=(k_1+k_2)/2$, and the Gaussian curvature $G_3=k_1 k_2$; $k_1$ and $k_2$ are the principal curvatures of $\partial K$, $dV$ and $dA$ are the infinitesimal volume and area element. In this work, the body $K$ is defined as the set of all points with overdensity larger than the threshold $\delta$. As can be seen from Eq~\ref{eq:rtos}, the velocity field introduces anisotropies to the overdensity field in redshift space. However, the integral~\ref{eq:w0} or~\ref{eq:wnu} collects contributions from the infinitesimal volume or area element in an angle-independent way, this means anisotropies introduced by RSD can be buried when adding up contributions that involve the LoS direction and those that involve directions orthogonal to the LoS \cite{2013MNRAS.435..531C,2021arXiv210803851J}. To help understand the lost information in the MFs, let's take an ellipsoid with semiaxes $(a,b,c)$ as a toy model for the excursion set. In real space, $a=b=c$, the ellipsoid becomes a spheroid. In redshift space, the longest semiaxis aligns along the LoS when the Fingers of God (FoG) effect \cite{1972MNRAS.156P...1J} dominates while its direction is perpendicular to the LoS when the Kaiser effect \cite{1987MNRAS.227....1K} dominates. The MFs can measure the change in the volume, surface area, and integrated mean curvature between the spheroid in real space and the ellipsoid in redshift space, but they can not tell the difference between the ellipsoid with the direction of its major axis parallel with or perpendicular to the LoS. This motivates us to explore their generalization to tensorial quantities, the Minkowski tensors.

The Minkowski tensors are generated by symmetric tensor products of position vectors $\mathbf{r}$ and normal vectors $\mathbf{n}$ of $\partial K$, those of rank 2 are defined as
\begin{equation}
	W^{2,0}_{0}(K)=\int_K \mathbf{r}^2dV,
\end{equation}
and
\begin{equation}
	W^{l,m}_{\nu}(K)=\frac{1}{3}\int_{\partial K}G_{\nu}\mathbf{r}^l\mathbf{n}^mdA,
\end{equation}
with $(l,m)=(2,0),\ (1,1)$ or $(0,2)$, the corresponding tensor (or dyadic) products are: $\mathbf{r}^2:=\mathbf{r} \otimes \mathbf{r}$, $\mathbf{r}\mathbf{n}:=\mathbf{r} \otimes \mathbf{n}$, and $\mathbf{n}^2:=\mathbf{n} \otimes \mathbf{n}$, respectively. The $(i,j)$ element of the tensor product is $(\mathbf{r} \otimes \mathbf{n})_{ij}:=\mathbf{r}_i \mathbf{n}_j$, where $i,\ j$ subscripts run over a Cartesian $(x_1,\ x_2,\ x_3)$ or equally $(x,y,z)$ coordinate system. In this work, we only study the translation invariant tensors $W^{0,2}_{1}$ and $W^{0,2}_{2}$ , because they remain unchanged under translations \cite{2013NJPh...15h3028S}, thus their measurement does not depend on the choice of coordinate origins.

The $(i,j)$ component of $W^{0,2}_1$ and $W^{0,2}_2$ is given as 
\begin{equation}
\label{eq:w102}
	(W^{0,2}_{1})_{i,j}=\frac{1}{3}\int_{\partial K}n_in_jdA, 
\end{equation}
and 
\begin{equation}
\label{eq:w202}
        (W^{0,2}_{2})_{i,j}=\frac{1}{3}\int_{\partial K}G_{2}n_in_jdA.
\end{equation}
 For an isotropic field, the diagonal components of MTs $(i,j)=(1,1),\ (2,2),\ (3,3)$ should be statistically the same, and the off-diagonal elements should be consistent with zero. While for a linearly redshift-space-distorted Gaussian random field, \cite{2019ApJ...887..128A} derived the ensemble expectation values for the perpendicular and parallel components of both $W^{0,2}_{1}$ and $W^{0,2}_{2}$. Take the prediction of $W^{0,2}_{1}$ as an example, 
\begin{equation}
\label{eq:w10211_lambda}
(W^{0,2}_{1})_{11} \propto \frac{\sigma_{1 \perp}}{\sigma}\left[\frac{\left(2 \lambda^2-1\right) \cosh ^{-1}\left(2 \lambda^2-1\right)}{\left(\lambda^2-1\right)^{3 / 2}}-\frac{2 \lambda}{\lambda^2-1}\right],
\end{equation}
while 
\begin{equation}
\label{eq:w10233_lambda}
(W^{0,2}_{1})_{33} \propto \frac{\sigma_{1 \perp}}{\sigma}\left(\frac{\lambda^2}{\lambda^2-1}\right)\left(\lambda-\frac{\cosh ^{-1} \lambda}{\sqrt{\lambda^2-1}}\right),
\end{equation}
and $\lambda = \sigma_{1 \|}^2 / \sigma_{1 \perp}^2$ is a function of the redshift space distortion parameter $\beta = f/b$. $\sigma_{1 \|}^2$ and $\sigma_{1 \perp}^2$ are the modified cumulants due to the effect of redshift-space distortion. \footnote{ The modified cumulants are defined as: $\sigma_{1,\perp} = \sqrt{1+\frac{6 \beta}{15}+\frac{3 \beta^2}{35}}\sigma_1$ and $\sigma_{1,\|} = \sqrt{1+\frac{6 \beta}{5}+\frac{3 \beta^2}{7}}\sigma_1$, where $\sigma_1^2=\frac{1}{2 \pi^2} \int d k k^{2+2 } P_{\mathrm{m}}(k)$ and $P_{\mathrm{m}}(k)$ is the matter power spectrum.} The two equations give us some insights into where the constraining power of the MTs on parameters related to RSD comes from. The non-Gaussian RSD effect on the perpendicular and parallel components of the MTs has not been investigated yet in the literature.

To measure the three-dimensional Minkowski tensors, we first apply redshift space distortion along the z-axis and then interpolate the positions of  CDM particles onto a grid with $N_{grid}=512$ using the cloud-in-cell (`CIC') \footnote{We use the routine provided by Pylians (Python libraries for the analysis of numerical simulations \cite{2018ascl.soft11008V}): https://pylians3.readthedocs.io/en/master/, other mass assignment schemes are also available, such as `NGP' (nearest grid point), `TSC' (triangular-shape cloud), and 'PCS' (piecewise-cubic-spline).} mass assignment scheme. The density field is then transformed to the density contrast field with $\delta (\boldsymbol{r})=(\rho(\boldsymbol{r})-\bar{\rho})/\bar{\rho}$. Hereafter, the word ``density'' will refer to density contrast most of the time. We smooth the density field via the Fourier-space convolution of
the density field with a Gaussian kernel
\begin{equation}
\label{eq:Gaussian_filter}
W(\vec{x}, R_G)=\frac{1}{(2 \pi)^{3 / 2} R_G^3} \exp \left(-\frac{\vec{x} \cdot \vec{x}}{2 R_G^2}\right)
\end{equation}
to suppress shot noise and increase the smoothness of the field using the routines provided by the package Pylians. Two different smoothing scales, $R_G=5,\ 10h^{-1}\rm{Mpc}$, are used to probe multi-scale structures. We then generate the bounding surface of an excursion set using the numerical algorithm described in Appendix of \cite{2018ApJ...863..200A}. Although the precision of statistics calculated based on the isodensity contours generated with this method is much higher than those based on the surface of voxels, there still exist discretization errors in these iso-density surfaces. This topic is also discussed in the same Appendix, where we can find in Figure 13 that the relative error in both $W_1^{0,2}$ and $W_2^{0,2}$ is smaller than $1\%$ when $R_G>2.5\ l_{grid}$. Since the grid size $l_{grid}=1000/512h^{-1}\rm{Mpc}$ in this work, both the two smoothing scales $R_G=5,\ 10h^{-1}\rm{Mpc}$ satisfy this criteria. We note there exist some intrinsic anisotropies in the algorithm used to generate the isodensity contours when the smoothing scale is not much larger than the grid size \footnote{We refer to Appendix B.3. of \cite{2018ApJ...863..200A} for a detailed discussion about this numerical issue, where they found the physical anisotropic signal would still dominate in the existence of this issue. }. This is why we find non-vanishing off-diagonal elements in the MTs for $R_G=5h^{-1}\rm{Mpc}$ when $l_{grid}=1000/512h^{-1}\rm{Mpc}$. A smaller $l_{grid}$ can alleviate this numerical issue but it demands so much more computational resources that we can't afford. Since our Fisher forecasts only use the diagonal elements, we anticipate that our results may not be significantly influenced by numerical issues existing in the measurement of the MTs. The Minkowski tensors of these isodensity surfaces are measured using the program karambola \cite{schroder2011minkowski,2013NJPh...15h3028S}. We have slightly modified it for the measurement of Minkowski tensors with the periodic boundary condition. 

We note that the spatial densities of both the Minkowski functionals and Minkowski tensors instead of themselves are used in this work, that is, we divide the four MFs and MTs with by volume of the simulation box. This is done by convention because it enables the direct comparison between the statistics measured for different volumes. Thus $W_{0}$ refers to the volume fraction, while $W_{1}$, $W_{2}$, $W_{3}$, and $W_{1,2}^{0,2}$ are the surface area, the integrated mean curvature, the Euler characteristic, and the Minkowski tensors per unit volume, respectively. 

\begin{figure}[tbp]
	\centering 
	\includegraphics[width=1.0\textwidth]{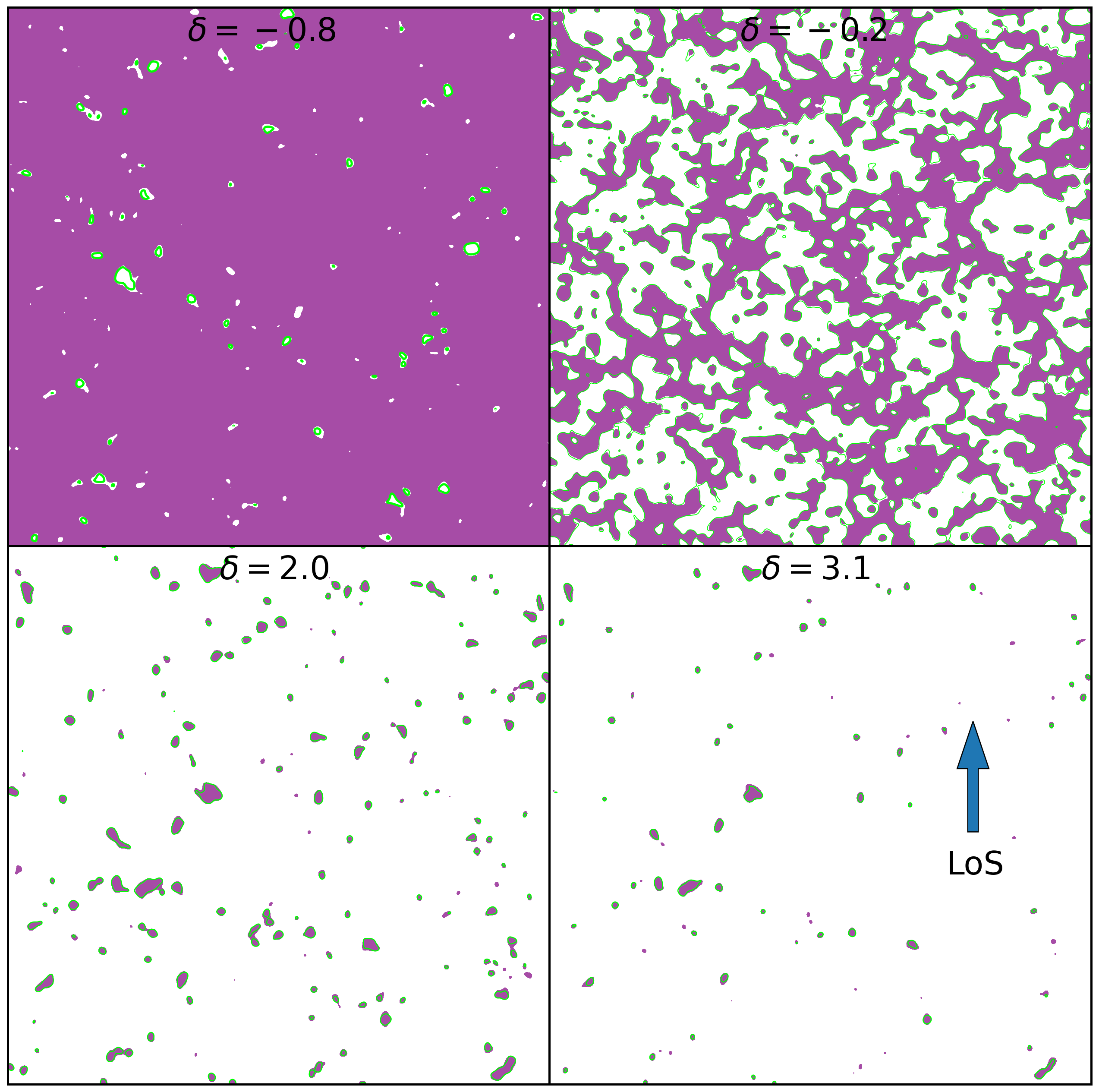}
	\caption{\label{fig:Contours} For 2D slices (thickness $2h^{-1}\rm Mpc$) of the redshift space 3D density field smoothed with $R_G=5h^{-1}\rm{Mpc}$, the 2D isodensity contours in a region of $1000\times 1000\ (h^{-1}\rm{Mpc})^2$ at the fiducial cosmology are plotted in green lines while the 2D excursion sets at the $f_R^{++++}$ cosmology are shown with purple regions. From the top left to bottom right, the four panels display plots for $\delta=-0.8,\ -0.2,\ 2.0,\ $ and 3.1, respectively. The dominant structure in the four panels is thus voids, filaments, regions around clusters, and clusters.}
\end{figure}

\begin{figure}[tbp]
	\centering
	\includegraphics[width=1.0\textwidth]{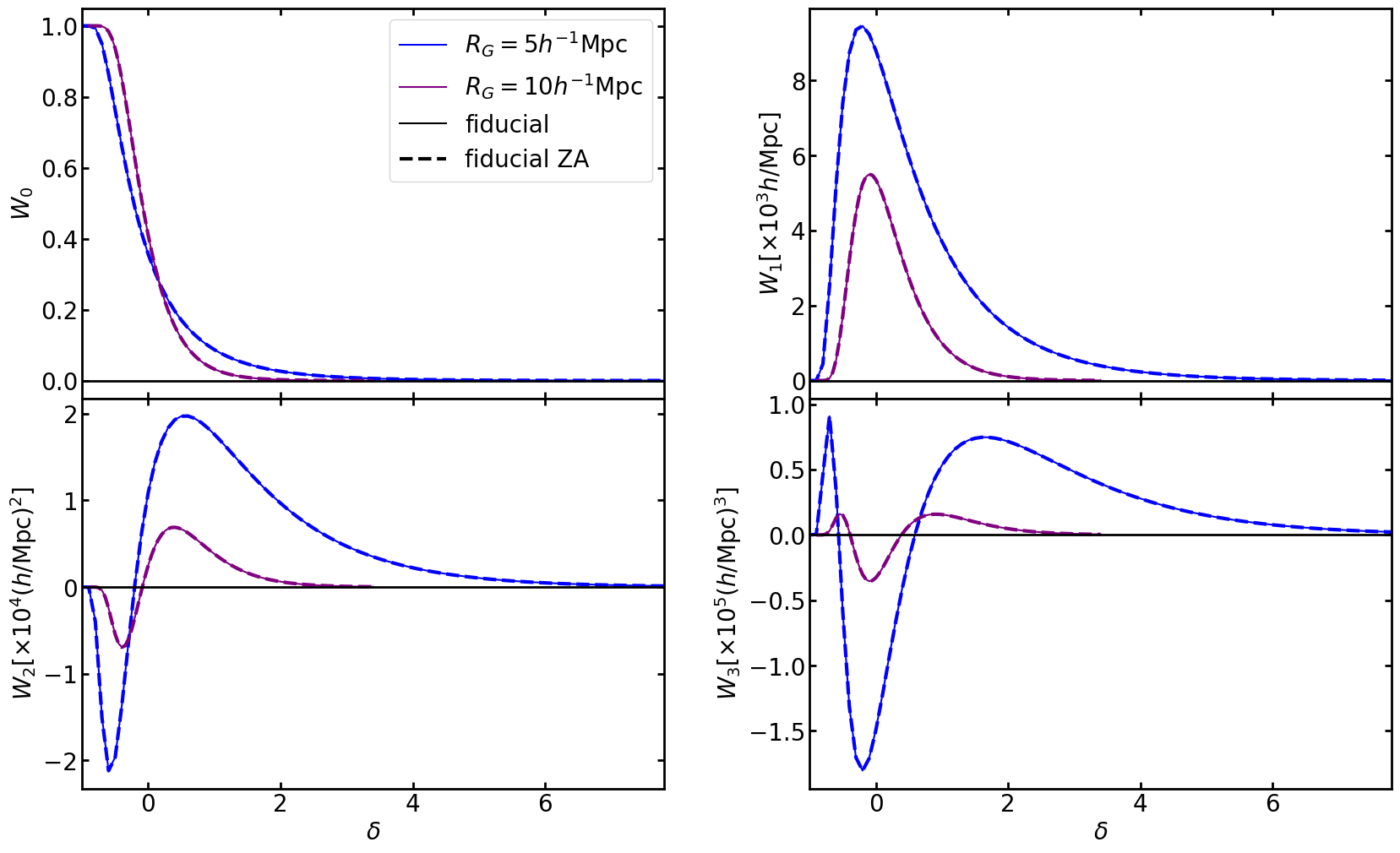}
	\caption{\label{fig:MFs_Rg5_10} The MFs for the fiducial simulations run with ICs generated using the second-order perturbation theory (solid lines, denoted as `fiducial') or Zel’dovich approximation (dashed lines, denoted as `fiducial ZA'). The MFs are measured as a function of the density contrast $\delta$ and averaged over 500 realizations for both the two different ICs. Blue lines for $R_G = 5h^{-1}\rm Mpc$ and purple lines for $R_G = 10h^{-1}\rm Mpc$.} 
\end{figure}

\begin{figure}[tbp]
	\centering 
	\includegraphics[width=1.0\textwidth]{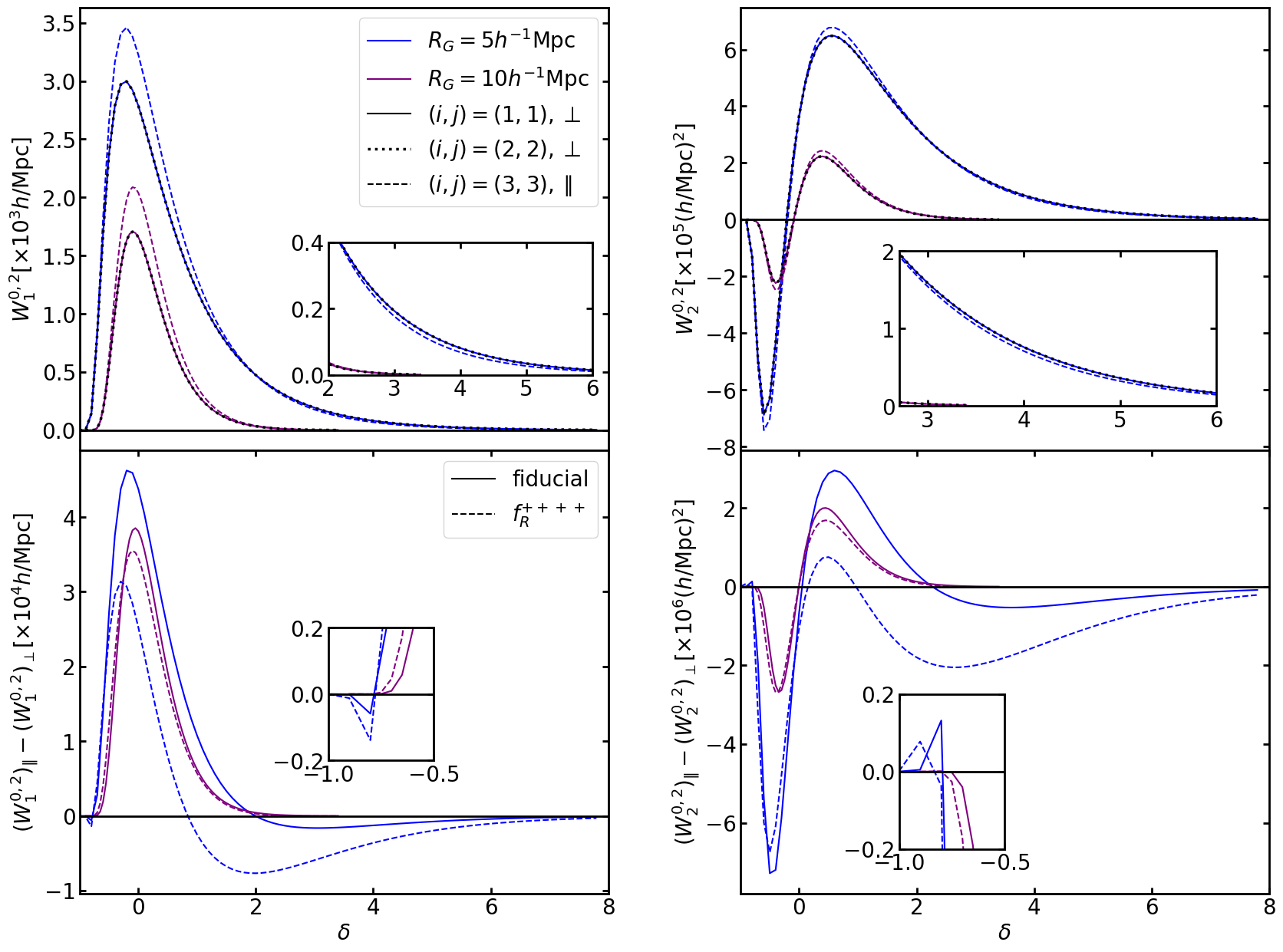}
	\caption{\label{fig:MTs} Top left panel: The perpendicular (solid and dotted lines, denoted as $(i,j)=(1,1),\ \perp$ and $(i,j)=(2,2),\ \perp$) and parallel component (dashed lines $(i,j)=(3,3),\ \parallel$) of $W_1^{0,2}$ are shown as functions of density threshold $\delta$. Blue for $R_G=5h^{-1}\rm{Mpc}$ and purple for $R_G=10h^{-1}\rm{Mpc}$. All curves in black are for $(i,j) = (2,2)$. The MTs are averaged over 5000 realizations at fiducial cosmology. Top right panel: Same as the top left panel, but for $W_2^{0,2}$. Bottom left panel: $(W_1^{0,2})_{\|}-(W_1^{0,2})_{\perp}$ for both fiducial (solid lines) and $f_R^{++++}$ (dashed lines) cosmology. The curves for $R_G=5,\ 10\ h^{-1}\rm{Mpc}$ are also plotted in blue and purple, respectively. Bottom right panel: Same as the bottom left panel, but for $W_2^{0,2}$.}
\end{figure}

To help understand the morphological and anisotropic information captured in the MTs, we plot in figure~\ref{fig:Contours} both the 2D isodensity contour at the fiducial cosmology and the excursion set \footnote{It is the set of all points whose density is higher than the density threshold, and is bound by the isodensity contour.} at the $f_R^{++++}$ cosmology
for $\delta=-0.8$ (top left), $\delta=-0.2$ (top right), $\delta=2.0$ (bottom left), and $\delta=3.1$ (bottom right). These contours and regions are constructed from the redshift space density field smoothed with $R_G=5h^{-1}\rm{Mpc}$ and RSD applied along the vertical direction. The 3D four MFs are shown in figure~\ref{fig:MFs_Rg5_10} for the fiducial simulations run with the ICs generated using the 2LPT or Zel'dovich approximation. Overall, the MFs for the two ICs agree well with each other, no matter whether the density field is smoothed with a Gaussian filter of $R_G=5h^{-1}\rm Mpc$ or $R_G=10h^{-1}\rm Mpc$. We find the different ICs only lead to differences slightly larger than the uncertainties of the MFs for a narrow threshold range around $\delta \sim -0.7$ for $R_G=5h^{-1}\rm Mpc$. Therefore, although the derivatives of the statistics (see Section~\ref{sec:fisher_deri}) w.r.t. $f_{R_0}$ and $M_{\nu}$ are calculated based on simulations with the Zel'dovich approximation ICs while the derivatives w.r.t. other parameters are obtained from simulations with the 2LPT ICs, we don't anticipate this difference in the ICs can significantly influence our results. In addition, when calculating the derivatives w.r.t. $f_{R_0}$ and $M_{\nu}$ in our Fisher forecasts, we take the difference between non-fiducial and fiducial simulations, which are both run with the same ICs. Therefore, even if the statistics themselves are impacted by the ICs, maybe their derivatives are not, as the systematic differences due to the ICs may cancel out.

The contours shown in figure~\ref{fig:Contours} display how the excursion set evolves as a function of $\delta$. As the thresholds increase, voids first appear, surrounded by the excursion sets. These voids grow in size until filaments emerge and dominate. The filaments then shrink, gradually revealing an increasing number of high-density islands. The islands contract into halos and eventually disappear. Since the total volume fraction of the voids is given by $1-W_0$ while that of the filaments and high-density islands is equal to $W_0$, $W_0$ is supposed to decrease when increasing $\delta$. $W_1$ measures the surface area of these large-scale structures, therefore, it should first increase and then decrease as a function of $\delta$. The mean curvature is negative for concave voids but positive for convex high-density regions. That is why $W_2$ is negative when $\delta$ is small and gradually becomes positive for large $\delta$. The transition from negative to positive happens when the structure is dominated by filaments with complicated shapes. $W_3$ is equal to the number density of isolated structures (like voids and islands) minus that of holes or tunnels which mainly exist with filaments. Therefore, $W_3$ is positive for low $\delta$ values, negative in the intermediate $\delta$ range, and positive when $\delta$ is large.

When larger smoothing scales are used, variations in the density field are reduced. That is, the threshold range where the MFs have non-vanishing values is shrunk. The surfaces of excursion sets become smoother, and small details that can increase the surface area are smeared out. The amplitude of $W_1$ and $W_2$ is thus smaller for $R_G=10h^{-1}\rm Mpc$ than that for $R_G=5h^{-1}\rm Mpc$. In addition, adjacent small voids, tunnels, and islands can be merged into one big void, tunnel, and island when larger smoothing scales are used, respectively. Hence the amplitude of $W_3$ is also smaller for $R_G=10h^{-1}\rm Mpc$. 

The 3D MTs $W_1^{0,2}$ and $W_2^{0,2}$ are plotted in the left and right columes of figure~\ref{fig:MTs}, respectively. The three components of MTs are displayed in the first row, while the difference between the parallel and perpendicular elements is shown in the second row for both the fiducial and $f_R^{++++}$ cosmology. The overall shape of $W_1^{0,2}$ and $W_2^{0,2}$ follows that of $W_1$ and $W_2$, respectively, which measure the surface area and the mean curvature integral on the surface of the excursion set. Since we have applied the redshift space distortions along the z-axis, the line-of-sight component $(i,j)=(3,3)$ is different from the perpendicular $(i,j)=(1,1),\ (2,2)$ components, this can be seen for both $W^{0,2}_{1}$ and $W^{0,2}_{2}$. In addition, since $W_{1,2}=(W^{0,2}_{1,2})_{1,1}+(W^{0,2}_{1,2})_{2,2}+(W^{0,2}_{1,2})_{3,3}$, there exists overlapped information in the MFs $W_{1,2}$ and the MTs $W^{0,2}_{1,2}$. Thus we will only perform Fisher matrix analysis for either $W_0+W_1+W_2+W_3$ or $W_0+W_1^{0,2}+W_2^{0,2}+W_3$. On the other hand, $(W^{0,2}_{1,2})_{1,1}$ is statistically equal to $(W^{0,2}_{1,2})_{2,2}$ and they are statistically independent, their average $(W^{0,2}_{1,2})_{\perp} = ((W^{0,2}_{1,2})_{1,1}+(W^{0,2}_{1,2})_{2,2})/2$ will have smaller uncertainty and will be used in the following Fisher matrix analysis, where the MTs will be composed of $(W_{1,2}^{0,2})_{\perp}= ((W^{0,2}_{1,2})_{1,1}+(W^{0,2}_{1,2})_{2,2})/2$ and $(W_{1,2}^{0,2})_{\|}=(W_{1,2}^{0,2})_{3,3}$.

We can compare the RSD effects on LSS between the fiducial and $f_R^{++++}$ cosmology in both figure~\ref{fig:Contours} and figure~\ref{fig:MTs}.
For $\delta=-0.8$, voids (or more precisely, central regions of voids) are delineated by the contours and excursion sets. A slightly larger number of voids in both the fiducial and $f_R^{++++}$ cosmology appear elongated along the line of sight (LoS) in redshift space. This means the normal vector for the infinitesimal area elements of voids is more likely to be perpendicular to the LoS. That is, the integrant $n_i n_j$ in Eq~\ref{eq:w102} for $i=j=1$ or $i=j=2$ is larger than for $i=j=3$ most of the time when integrating over the surface of voids. On the other hand, the sign of $W_2^{0,2}$ is negative for voids. Therefore, $(W_1^{0,2})_{\|}-(W_1^{0,2})_{\perp}<0$ and $(W_2^{0,2})_{\|}-(W_2^{0,2})_{\perp}>0$ \footnote{We find the sharp features observed in the bottom panels of figure~\ref{fig:MTs} become smooth when more threshold bins are added in the range $-1.0$ to $-0.5$. They appear sharp because only a single data point exists in that region, and we do not apply interpolation or curve smoothing when plotting the MT curves.}, and can be seen in figure~\ref{fig:MTs}, although it is a very small signal. We can understand why voids are elongated in redshift space: since the surrounding CDM particles are moving out of these low-density regions; the particles farther from us are moving away from us, so they appear farther than they actually are; similarly, particles close to us are moving toward us, so they appear closer than they actually are \cite{2021MNRAS.500..911C,2012arXiv1203.0869S}.

For $\delta=-0.2$, the cosmic structure is dominated by complicated filaments
and it is thus hard to find out whether the filaments are squashed due to the Kaiser effect \cite{1987MNRAS.227....1K} or not. However, the bottom left panel of figure~\ref{fig:MTs} tells us these structures are less flattened in the $f_R^{++++}$ than those in the fiducial cosmology. Many supporting examples can be easily found in the top right panel of figure~\ref{fig:Contours}, where the 2D isodensity contours at the fiducial cosmology appear more flattened along the LoS than the excursion sets at the $f_R^{++++}$ cosmology.
 
The Kaiser and FoG effects are competing against each other for intermediate density thresholds. For $\delta=2.0$, the FoG effect has already dominated on LSS for the $f_R^{++++}$ cosmology, while the two effects are equally strong in the fiducial cosmology. Therefore, we can see most of the high-density regions shown in the bottom left panel of figure~\ref{fig:Contours} are elongated along the LoS in the $f_R^{++++}$ cosmology. However, these structures are either less stretched along the LoS or split into smaller and more isotropic islands in the fiducial cosmology. For $\delta=2.0$, the FoG effect finally wins and dominates in both the $f_R^{++++}$ and fiducial cosmology. And we can see the high-density regions are more elongated for $f_R^{++++}$ cosmology, this is because the fifth force makes the particles move faster, increasing the velocity dispersion and causing a stronger FoG effect compared to GR \cite{Jennings:2012pt}.

\section{Fisher matrix formalism}
\label{sec:fisher} 
We use the Fisher information matrix \cite{fisher1936use,tegmark1997karhunen} to quantify the constraining power of the statistics considered in this work on the cosmological parameters, which is defined as
\begin{equation}
	F_{\alpha \beta}=\left\langle-\frac{\partial^{2} \ln \mathcal{L}}{\partial \theta_{\alpha} \partial \theta_{\beta}}\right\rangle,
\end{equation}
where the likelihood $\mathcal{L}$ is assumed to be Gaussian. We will check in Appendix \ref{sec:Gaussian_test} the likelihood for all statistics and their combinations studied in this work can indeed be approximated by Gaussian. To avoid introducing artificial information, the covariance matrix of the observables $\boldsymbol{C}$ is assumed to be parameter-independent \cite{2013A&A...551A..88C}. The Fisher matrix can then be written as
\begin{equation}
\label{eq:fisher_mat}
	F^{stand}_{\alpha \beta}=\frac{\partial \boldsymbol{\mu}}{\partial \theta_{\alpha}}^{\mathrm{T}} \boldsymbol{C}^{-1} \frac{\partial \boldsymbol{\mu}}{\partial \theta_{\beta}},
\end{equation}
where $\boldsymbol{\mu}$ is the theoretical mean for the data vector. It can be the power spectrum, MFs, MTs, and the combination of these statistics. $\boldsymbol{C}^{-1}$ is the inverse of the covariance matrix. To differentiate this estimator from the other two estimators that will be introduced in the following, we will call it the standard estimator of the Fisher matrix.

Recently, \cite{2024arXiv240606067W} finds the statistical noise in the numerical derivatives of observables can bias the Fisher forecast and underestimate the marginalized errors, the bias will be larger when higher-order forward differentiation schemes are used to estimate derivatives w.r.t. parameters like $M_{\nu}$. However, if we first compress the observables with the optimal compression scheme \cite{2018MNRAS.476L..60A} 
\begin{equation}
\label{eq:t_comp}
	t_{\alpha}=\boldsymbol{\mu}_{\alpha}^{\mathrm{T}} \boldsymbol{C}^{-1}  (\boldsymbol{d}-\boldsymbol{\mu}),\   \boldsymbol{\mu}_{\alpha}\equiv \frac{\partial \boldsymbol{\mu}}{\partial \theta_{\alpha}},
\end{equation}
where $\boldsymbol{d}$ is the observable measured from a realization of simulations, and calculate the Fisher matrix for the compressed observable using the same equation as Eq~\ref{eq:fisher_mat}, but replace the derivatives and covariance matrix with those for the compressed observable, then the Fisher matrix will be oppositely biased and overestimate the marginalized errors \cite{2023arXiv230508994C}. This new estimator is denoted with $F^{comp}$ and we will call it the compressed estimator hereafter, we will follow \cite{2023arXiv230508994C} and combine it with the standard estimator to give less biased (if not unbiased) forecasts
\begin{equation}
\label{eq:F_comb}
	F^{comb}_{\alpha \beta}=G(F^{stand}_{\alpha \beta},F^{comp}_{\alpha \beta}), 
\end{equation}
where $G(A, B)$ is the geometric mean of matrix A and B,
defined as
\begin{equation}
\label{eq:GAB}
	G(A,B)=A^{1/2}(A^{-1/2}BA^{-1/2})^{1/2}A^{1/2}.
\end{equation}
Our Fisher matrix forecasts are all calculated with the combined estimator \footnote{We refer to \cite{2023arXiv230508994C} for a more detailed description of how to calculate the combined estimator and a Python package implementing these methods is available at https://github.com/wcoulton/CompressedFisher}, we will present a comparison of the forecasts from the standard, compressed, combined estimator and use it as an additional convergence test in Appendix~\ref{sec:conver}.

\subsection{Derivatives}
\label{sec:fisher_deri}

For cosmological parameters, except the $f(R)$ parameter $f_{R_0}$ and the sum of neutrino masses $M_{\nu}$, the derivatives are estimated as
\begin{equation}
\label{eq:para_deri}
	\frac{\partial \boldsymbol{\mu}}{\partial \theta_{\alpha}}\simeq\frac{\boldsymbol{\mu}(\theta_{\alpha}^{+}) - \boldsymbol{\mu}(\theta_{\alpha}^{-})}{\theta_{\alpha}^{+} - \theta_{\alpha}^{-}},
\end{equation}
where $\boldsymbol{\mu}(\theta_{\alpha}^{+})$ and $\boldsymbol{\mu}(\theta_{\alpha}^{-})$ are estimated as the average of observables from 500 realizations at $\theta_{\alpha}^{+}$ and $\theta_{\alpha}^{-}$, respectively.  Since the values of $f_{R_0}$ are not distributed equally in both linear and log, we first perform the following change of variables: $f_{R_0}$ to $|f_{R_0}|^{\lg 2}$, so that the values of $f_{R_0} = 0,\ -5\times10^{-7},\ -5\times10^{-6},\ -5\times10^{-5},\ -5\times10^{-4}$ map to $|f_{R_0}|^{\lg 2}=0,\ 0.0127,\ 0.0254,\ 0.0507,\ 0.101$. $f_{R_0}$ will still be used to refer to $|f_{R_0}|^{\lg 2}$ hereafter. Then the derivatives w.r.t. $f_{R_0}$ is estimated as
\begin{equation}
\label{eq:fR_deri}
	\frac{\partial \boldsymbol{\mu}}{\partial f_{R_0}}=\frac{\boldsymbol{\mu}(8df_{R_0})-12\boldsymbol{\mu}(4df_{R_0})+32\boldsymbol{\mu}(2df_{R_0})- 21\boldsymbol{\mu}(\theta^{ZA}_{fid})}{12df_{R_0}},
\end{equation}
where $df_{R_0}=0.0127$, and $\boldsymbol{\mu}(8df_{R_0})$, $\boldsymbol{\mu}(4df_{R_0})$, $\boldsymbol{\mu}(2df_{R_0})$, and $\boldsymbol{\mu}(\theta^{ZA}_{fid})$ are estimated with 500 realizations for model $f_R^{++++}$, $f_R^{+++}$, $f_R^{++}$, and Fiducial ZA, respectively. We don't use simulations for $f_R^{+}$ because it is so close to fiducial cosmology that the difference in statistics between $f_R^{+}$ and GR will be dominated by noise \cite{2024arXiv240718647V}. For $M_{\nu}$, the derivative is constructed with
\begin{equation}
\label{eq:nu_deri}
	\frac{\partial \boldsymbol{\mu}}{\partial M_{\nu}}=\frac{\boldsymbol{\mu}\left(4 d M_\nu\right) - \boldsymbol{\mu}\left(\theta_{fid}^{ZA}\right)}{4 d M_\nu},
\end{equation}
where $dM_{\nu}=0.1$eV, $\boldsymbol{\mu}(4dM_{\nu})$ and $\boldsymbol{\mu}(\theta_{fid}^{ZA})$ are estimated as the average of observables from 500 realizations at the $M_{\nu}=0.4eV$ cosmologies, and the fiducial cosmology with Zeldovich ICs, respectively. We choose this estimator because the noise in it is smaller than a similar estimator to Eq~\ref{eq:fR_deri}, which will give us a more convergent Fisher forecast \cite{2024arXiv240606067W}. We will discuss this in detail in Appendix~\ref{sec:conver}.

\begin{figure}[tbp]
	\centering 
	\includegraphics[width=1.0\textwidth]{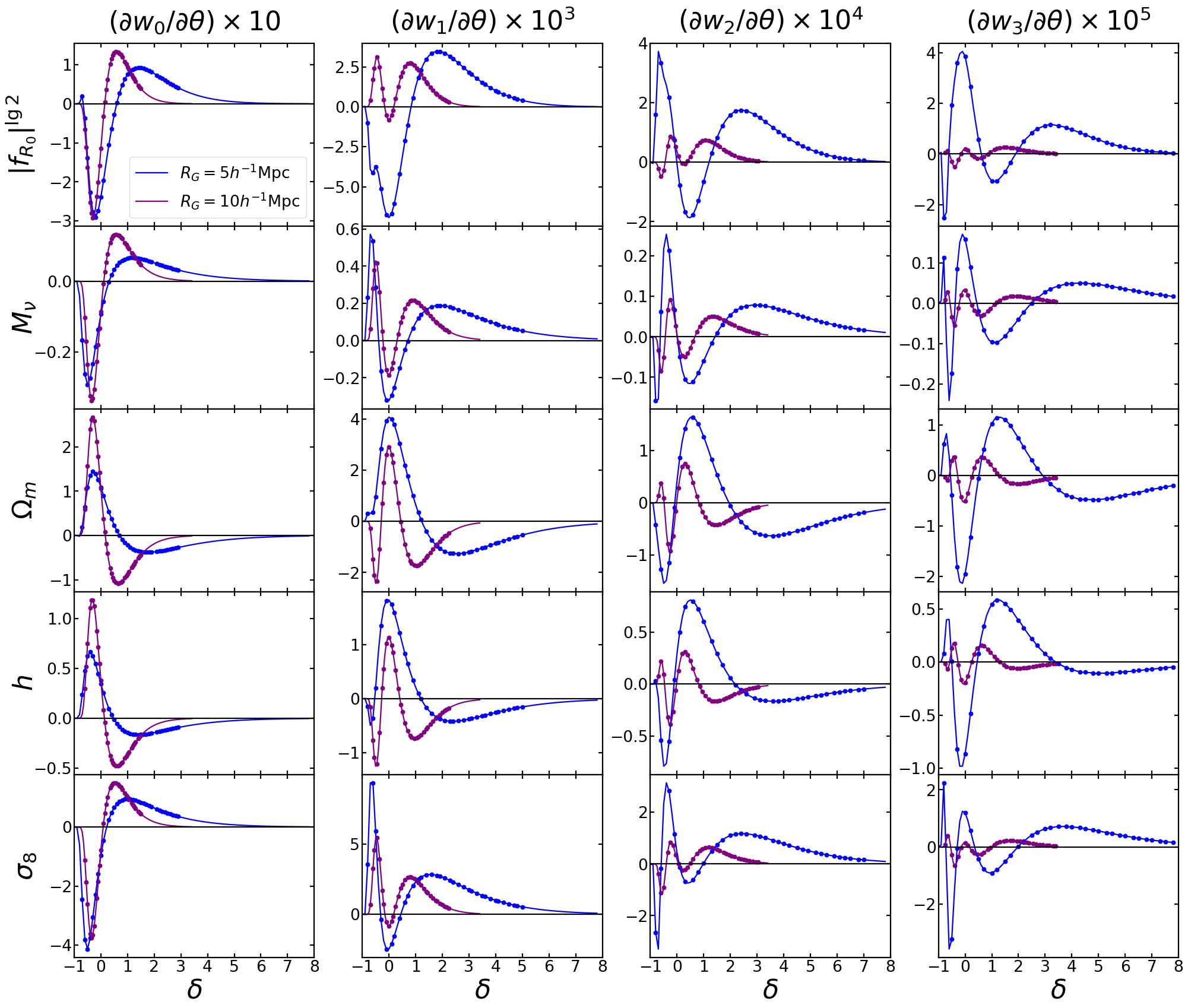}
	\caption{\label{fig:MF_deri} The numerical derivatives of the four 3D MFs w.r.t. the five cosmological parameters. The curves for $R_G=5,\ 10 h^{-1}\rm{Mpc}$ are plotted in blue and purple. The dots plotted on the curves denote the data points that will be used in the Fisher forecast.}
\end{figure}

\begin{figure}[tbp]
	\centering 
	\includegraphics[width=1.0\textwidth]{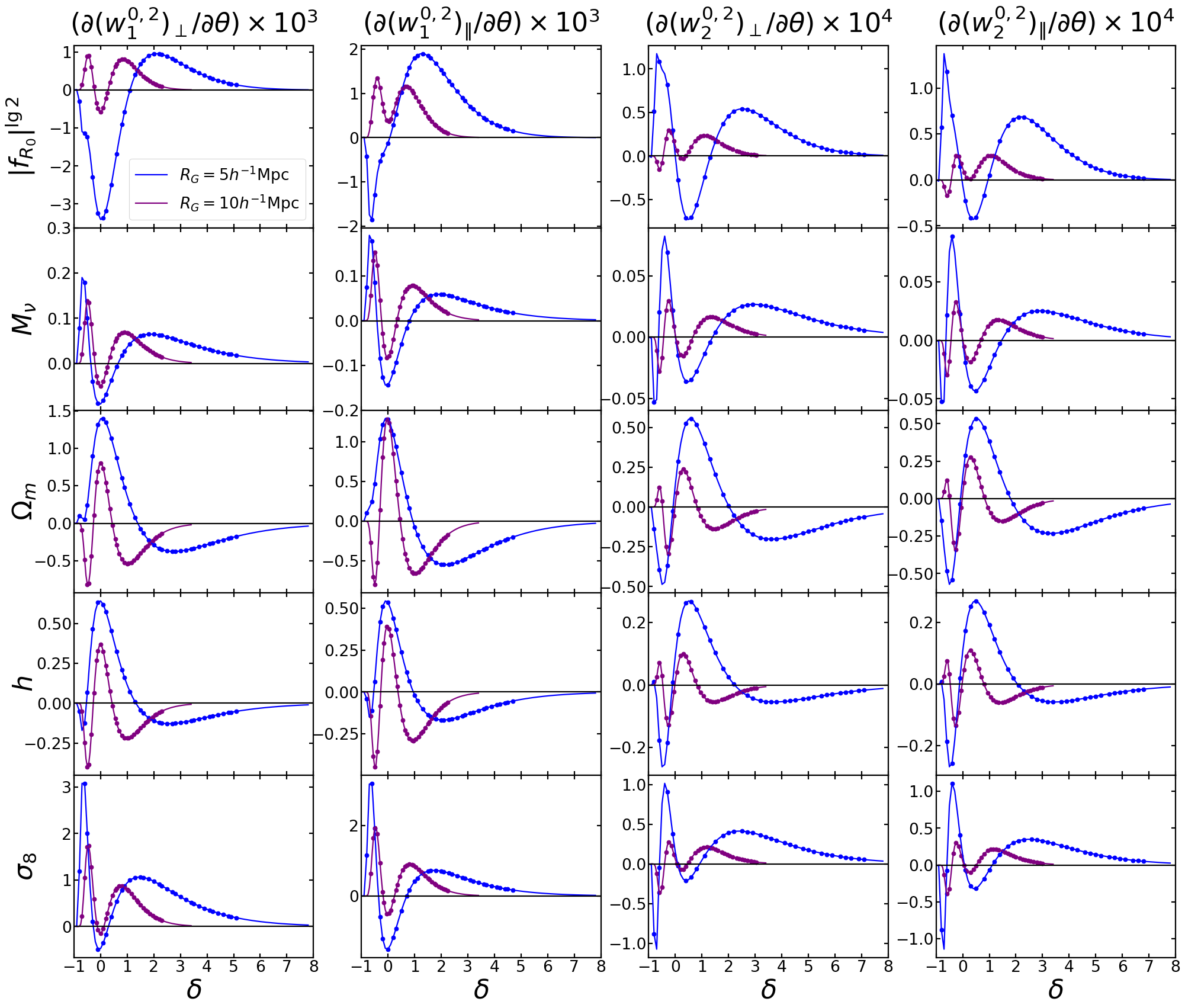}
	\caption{\label{fig:MT_deri} The numerical derivatives of the $(1,1)$ and $(3,3)$ components of the two 3D MTs $W_1^{0,2}$ (the first two columns) and $W_2^{0,2}$ (the last two columns) w.r.t. the five cosmological parameters. The curves for $R_G=5,\ 10h^{-1}\rm{Mpc}$ are plotted in blue and purple. The dots plotted on the curves denote the data points that will be used in the Fisher forecast. We note there also exists an evident difference between the $\perp$ and $\|$ components of both $\partial W^{0,2}_1/\partial \theta$ and $\partial W^{0,2}_2/\partial \theta$ for $|f_{R_0}|^{\lg 2}$.}
\end{figure}

For the data vector of the MFs and MTs, the thresholds are chosen based on the average statistics of 5000 fiducial simulations with the same binning scheme as that used in our previous work \cite{2023JCAP...09..037L}: $N_b$ evenly spaced threshold bins are taken for $W_0$ in the range between $\delta_{0.01}$ to $\delta_{0.99}$, where the thresholds $\delta_{0.01}$ and $\delta_{0.99}$ correspond to the area (volume) fraction of 0.01 and 0.99, respectively. For other MFs and MTs, we first find the lower end of thresholds approximately corresponding to $1\%$ of the maximum of the statistics, $\delta_{low}$; and then find the higher end of thresholds where the curve is close to $1\%$ of the maximum of the statistics as well, $\delta_{high}$. Finally, $N_b$ evenly spaced threshold bins are taken in the range between $\delta_{low}$ to $\delta_{high}$. This threshold binning scheme tries to cover the variation range for each order of the statistics as extensively as possible while avoiding the very low and high thresholds where the variance of the MTs is zero and the likelihood of the MTs is non-Gaussian. The zero-valued variance will lead to numerical issues when calculating the Fisher matrix while the non-Gaussian likelihood may result in overestimated constraints from the MTs \cite{2022arXiv220405435P}.

For each of the monopole, quadrupole, and hexadecapole of the power spectrum, 80 wavenumber bins are used, up to $k_{\rm max}=0.5~h\rm{Mpc}^{-1}$. The size of each bin is $2 \pi/L$, where $L=1h^{-1} \rm Gpc$ is the size of the simulation box. 

In figure~\ref{fig:MF_deri}, we show the derivatives of the MFs w.r.t. the five cosmological parameters for two different smoothing scales $R_G=$ 5 (blue), 10 (purple) $h^{-1}\rm{Mpc}$. The data points chosen using the threshold binning scheme described above are marked with dots. It seems the curves in the low-density range are sparsely sampled with these points, but we find there is only a small change in the Fisher forecast when we increase the number of points in the low-density range. Since we explained in detail how the derivatives of the MFs change with the smoothing scale $R_G$ in \cite{2023JCAP...09..037L}, we will not repeat it here and only point out there are less distinct features in the derivatives for $R_G=10h^{-1}\rm{Mpc}$ than $R_G=5h^{-1}\rm{Mpc}$ because important structure information is smeared out when using the larger smoothing scale. Focusing on the curves for $R_G=5h^{-1}\rm{Mpc}$, we find that $\partial W_i/\partial \theta$ for $|f_{R_0}|^{\lg 2}$ is quite different from that for $M_{\nu}$ and the difference is most conspicuous for low-density thresholds: compared to $\partial W_i/\partial \theta$ for $M_{\nu}$, there is an extra peak on $\partial W_0/\partial \theta$ for $f_{R_0}$; a small wiggle rather than a peak lies on the left slope of the large valley of $\partial W_1/\partial \theta$; a small valley and a small peak are missing for $\partial W_2/\partial \theta$ and $\partial W_3/\partial \theta$, respectively. The overall shape of $\partial W_i/\partial \theta$ for $\Omega_m$ is close to that for $h$, but the difference in $\partial W_1/\partial \theta$ for low-density threshold can be easily detected by human eyes. The derivatives for $M_{\nu}$ and $\sigma_8$ are most similar, only varied height, depth, width, and position of peaks and valleys can be found by careful comparison. The strong degeneracy between $M_{\nu}$ and $\sigma_8$ will be broken when information from the total matter field is included \cite{2022JCAP...07..045L,2021PhRvL.126a1301M,2021arXiv210807821V,2021ApJ...919...24B}. This degeneracy exists in both MFs and MTs for the CDM particle field, more convergent Fisher forecast results will be obtained if it is broken, as will be discussed in Appendix~\ref{sec:conver}.

Besides the morphological information captured in the MFs, we also plot the derivatives of the perpendicular and parallel components of $W^{0,2}_1$ and $W^{0,2}_2$ in figure~\ref{fig:MT_deri}, where we can directly see the anisotropic information contained in the MTs. The most important feature is a clear difference in $\partial (W^{0,2}_{1,2})_{\perp} / \partial \theta$ and $\partial (W^{0,2}_{1,2})_{\|} / \partial \theta$. The difference is largest for the derivative w.r.t. $|f_{R_0}|^{\lg 2}$, and second largest for $\Omega_m$, this is reasonable because the two parameters are most directly related to the growth rate of structure $f$, which is the MTs are sensitive to. The difference also means distinct information is embedded in the perpendicular and parallel elements of the MTs. We can expect that the parameter degeneracies will be broken by the combination of the $(W^{0,2}_{1,2})_{\perp}$ and $(W^{0,2}_{1,2})_{\|}$.

Similar to what has been found in the MFs, there are many characteristic signals in the derivatives of the MTs w.r.t. $|f_{R_0}|^{\lg 2}$, which are distinguished from those w.r.t. $M_{\nu}$. The curve for $\Omega_m$ shares a similar shape with that for $h$ in the high-density threshold range, but the difference in the derivatives for the very low-density range breaks the degeneracy between the two parameters. Unfortunately, this feature is smeared out when $R_G=10h^{-1}\rm{Mpc}$, which leads to a larger $\Omega_m-h$ degeneracy. We can find a consistent enhancement of the $\Omega_m-h$ degeneracy in the MTs from $R_G=5h^{-1}\rm{Mpc}$ to $R_G=10h^{-1}\rm{Mpc}$ in the contours plotted in Section~\ref{sec:pkvsmt}. The strong degeneracy between $M_{\nu}$ and $\sigma_8$ still exists in the MTs, their derivatives only differ in the position, width, and amplitude of peaks and valleys on the curves. 

\subsection{Covariance matrix}
\label{sec:fisher_cov}
We estimate the covariance matrices with 5000 fiducial simulations, which are found to be enough for converged parameter constraints in Appendix~\ref{sec:conver}. Due to uncertainties in the estimated covariance matrix $\hat{\mathbf{C}}$, the inverse of $\hat{\mathbf{C}}$ is not an unbiased estimator for $\mathbf{C}^{-1}$. Following \cite{hartlap2007your}, we remove the bias in $\hat{\mathbf{C}}^{-1}$ by 
\begin{equation}
	\label{eq:debiase}
	\mathbf{C}^{-1}=\frac{n-p-2}{n-1} \hat{\mathbf{C}}^{-1},
\end{equation}
where p is the number of observables in the data vector, and n is the number of simulations used to estimate  $\mathbf{C}$.

\begin{figure}[tbp]
	\centering 
	\includegraphics[width=1.0\textwidth]{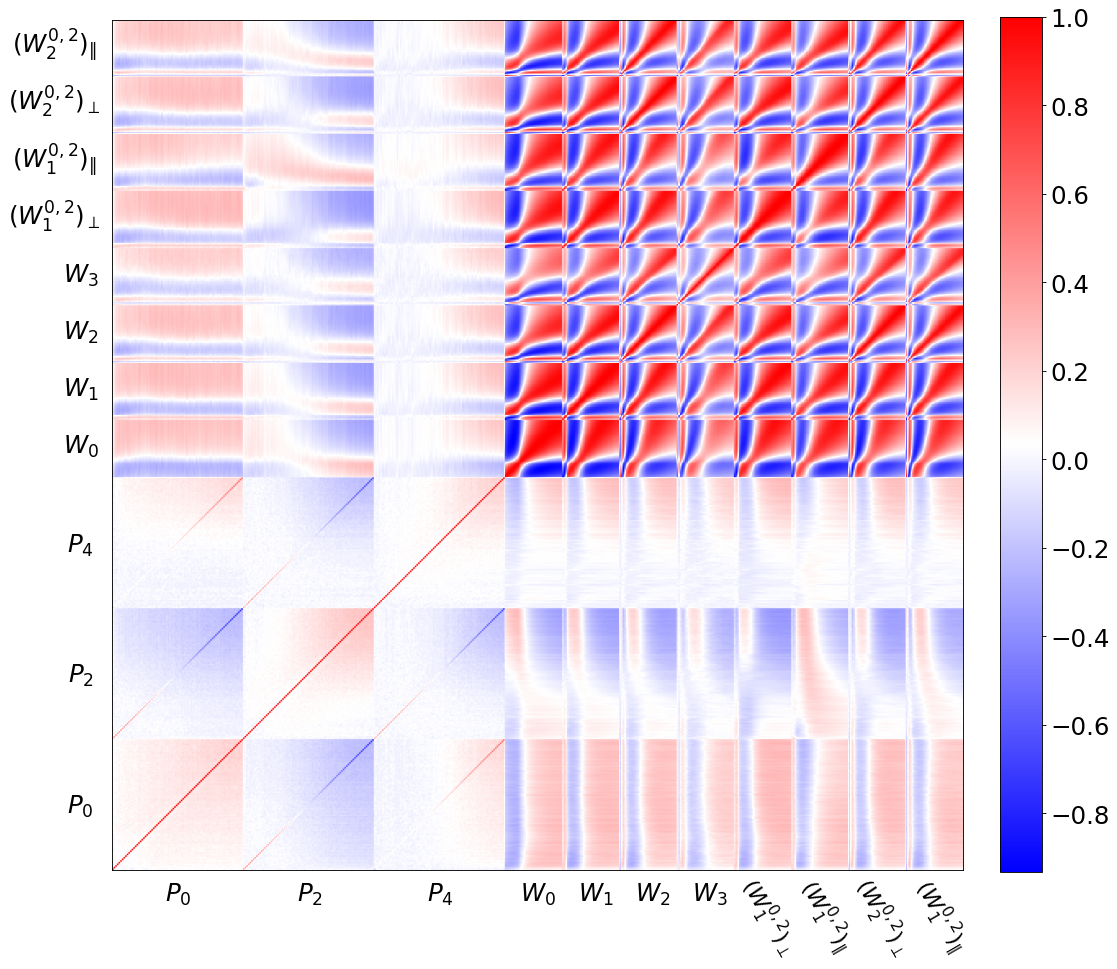}
	\caption{\label{fig:iv} Correlation matrix of the monopole, quadrupole, and hexadecapole of the power spectrum ($P_0,\ P_2,\ P_4$), MFs ($W_0,\ W_1,\ W_2,\ W_3$), and MTs ($W^{0,2}_{1},\ W^{0,2}_{2}$). For $P_0,\ P_2,\ P_4$, 80 wavenumber bins are used, up to $k_{\rm {max}}=0.5~h\mathrm{Mpc}^{-1}$. We only show the correlation matrix for the MFs and MTs with $R_G=5\ h^{-1}\rm Mpc$ here because the one for the MFs and MTs with $R_G=10\ h^{-1}\rm Mpc$ looks very similar. 35 threshold bins are taken for each order of MFs and each element of MTs. Bin values increase from left to right for each statistic. The correlation matrix is estimated using 5000 realizations of the fiducial model.}
\end{figure}

In figure~\ref{fig:iv}, we show the correlation matrices of the data vector that combines the power spectrum ($P_0+P_2+P_4$), MFs ($W_0,\ W_1,\ W_2,\ W_3$), and MTs ($W_1^{0,2},W_2^{0,2}$). For the power spectrum, 80 wavenumbers are used for each multipole, up to $k_{\rm max} = 0.5 h\ Mpc^{-1}$. For the MFs and MTs, 35 threshold bins are taken for each order of $W_i$, and each of $(W^{0,2}_{1})_{\perp}$, $(W^{0,2}_{1})_{\|}$, $(W^{0,2}_{2})_{\perp}$, $(W^{0,2}_{2})_{\|}$. 
Because there exists overlapped information in the MFs $W_{1,2}$ and the MTs $W^{0,2}_{1,2}$, we will only perform Fisher matrix analysis for either $W_0+W_1+W_2+W_3$ or $W_0+W_1^{0,2}+W_2^{0,2}+W_3$, a total of 140 or 210 bins will be used, respectively.

The correlation matrix of the power spectrum is shown in the left-bottom corner of figure~\ref{fig:iv}. The auto-correlation of each multipole of the power spectrum has a simple structure: cross-correlation between different wavenumbers is weak on large scales (small $k$-bins, bottom left), but gradually strengthens on smaller scales. We can see a similar cross-correlation between the monopole and quadrupole to that between the quadrupole and the hexadecapole: the correlation is strong between multipoles in the same $k$-bin, and it is positive for small $k$-bins but negative for large $k$-bins; the correlation is weak and mainly negative between multipoles in varied $k$-bins. A weak cross-correlation can be seen between the monopole and the hexadecapole, it is close to zero for small $k$-bins and becomes positive for large $k$-bins. The small overall correlations between the power spectrum and the Minkowski statistics indicate that the Minkowski statistics can provide complementary information. The correlation matrix of the Minkowski statistics exhibits a much richer structure. The correlations between $W_{1,2}$ and $W_{1,2}^{0,2}$ as well as between perpendicular and parallel components of $W_{1,2}^{0,2}$ can be easily understood according to Eq~\ref{eq:w102} and Eq~\ref{eq:w202} ; while the correlation between $W_{1}^{0,2}$ and $W_{2}^{0,2}$ is quite similar to that between $W_1$ and $W_2$. Other correlations between the MFs with different threshold bins and orders are described in detail in our previous works \cite{2022JCAP...07..045L,2023JCAP...09..037L}, thus are not given here.

When the Fisher matrix is calculated using the simulation-based method, a large number of simulations are needed for an accurate estimate of both the derivatives and covariance matrix to obtain converged parameter constraints \cite{2021arXiv210100298B,2020JCAP...03..040H,2022arXiv220601619C}. In Appendix~\ref{sec:conver}, we will discuss this issue in more detail.

\section{Results}
\label{sec:result}
We will first study how the anisotropic information is gathered in the Minkowski tensors in section~\ref{sec:info_MTs}, then investigate how the anisotropies captured by quadrupole and hexadecapole ($P_{2,4}$) of the power spectrum as well as by the MTs $W^{0,2}_{1,2}$ help break parameter degeneracies and tighten constraints in section~\ref{sec:iso_aniso}, finally compare the power spectrum with the MFs plus MTs and discuss how their combination can improve constraints on modified gravity, massive neutrinos, and other parameters in section~\ref{sec:pkvsmt}.

\subsection{Information in the Minkowski tensors}
\label{sec:info_MTs}

\begin{figure}[tbp]
	\centering 
	\includegraphics[width=1.0\textwidth]{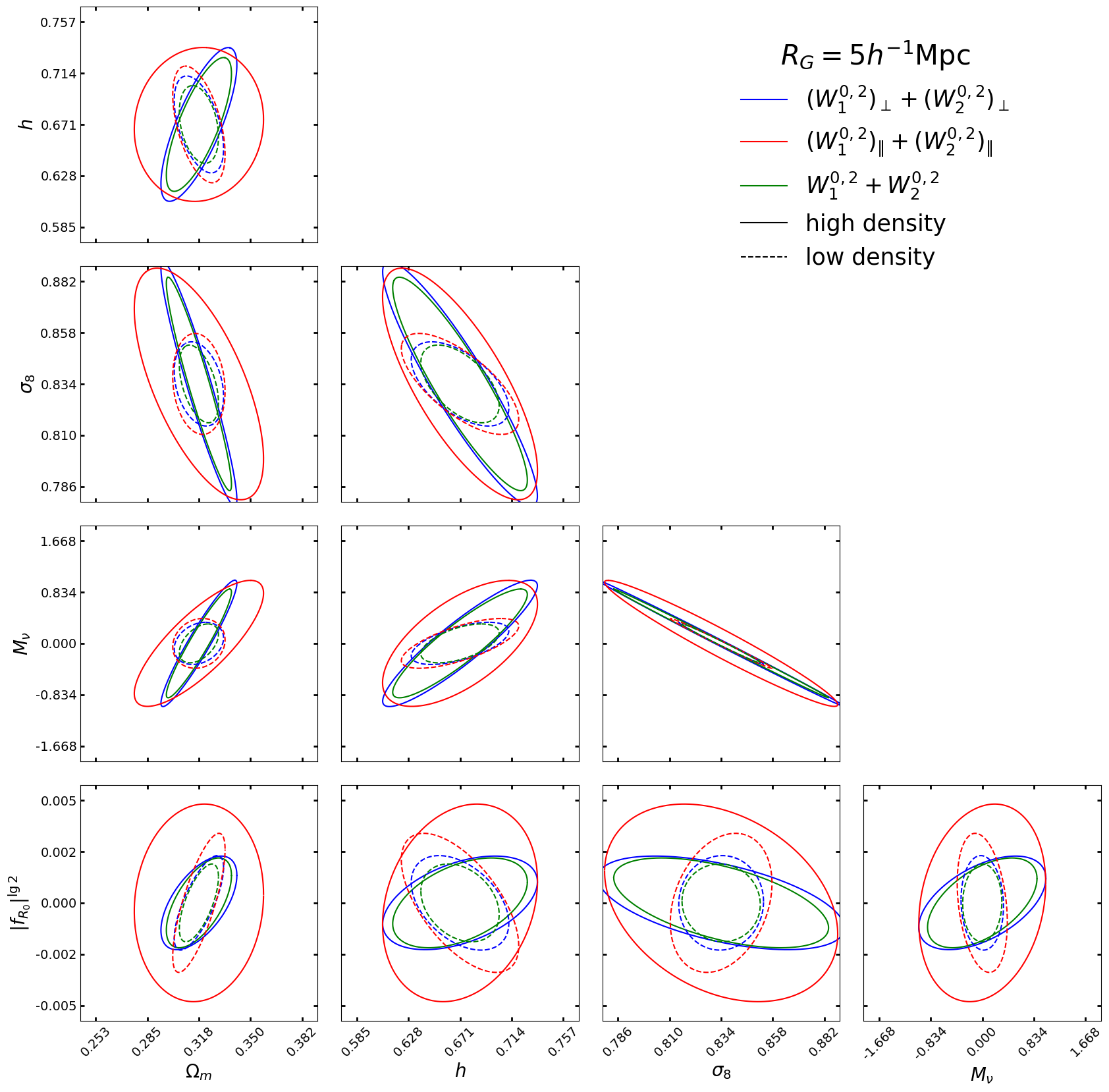}
	\caption{\label{fig:perp_plus_para} Parameter constraints for $R_G=5h^{-1}\rm{Mpc}$ from the perpendicular ($(W_{1}^{0,2})_{\perp}+(W_{2}^{0,2})_{\perp}$, blue), parallel ($(W_{1}^{0,2})_{\|}+(W_{2}^{0,2})_{\|}$, red) component of MTs, and their combination ($W_{1}^{0,2}+W_{2}^{0,2}$, green). The contours for high- and low-density thresholds are plotted with solid and dashed lines.}
\end{figure}

In section~\ref{sec:3dmts}, we have discussed the effect of RSD on isodensity contours for different thresholds. Three kinds of characteristic imprints of RSD on LSS are seen for $R_G=5h^{-1}\rm{Mpc}$: a slightly larger number of voids are elongated along LoS for very low-density thresholds; the more complicated contours are squashed along LoS due to the Kaiser effect for intermediate thresholds; and the FoG effect is seen for large thresholds where the high-density regions are also elongated along LoS. In contrast, the LSS is dominated by the Kaiser effect in the whole threshold range for $R_G=10h^{-1}\rm{Mpc}$. Therefore, we will only present results for $R_G=5h^{-1}\rm{Mpc}$ in this section, because we want to compare the information content of the MTs dominated by the Kaiser effect with that dominated by the FoG effect \footnote{We merge the first two kinds of RSD effect because we can see the elongated voids by RSD only for first two threshold bins, the Fisher forecasts will be dominated by the threshold bins where the Kaiser effect dominates}. The Kaiser effect captured by the MTs on linear scales is studied in \cite{2019ApJ...887..128A,2022arXiv220810164A}, where they found it is sensitive to the linear growth rate of the LSS, as shown in Eq~\ref{eq:w10211_lambda} and Eq~\ref{eq:w10233_lambda}. The information content in the MTs from the small-scale FoG effect is investigated and compared with that from the Kaiser effect in this work for the first time. We note that the information from the small-scale FoG effect is discussed in \cite{2020ApJ...897...17T} with the 2-point correlation function. The RSD effect in different density environments is also studied in \cite{2021MNRAS.505.5731P}.

For a fair comparison, 15 evenly spaced threshold bins are taken in the range between $\delta_{low}$ and $\delta = 2$ for ``low-density thresholds'', while 15 threshold bins are also evenly taken from $\delta > 2$ to $\delta_{high}$ for ``high-density thresholds''. $\delta_{low}$ and $\delta_{high}$ is the same as the threshold binning scheme described in Section~\ref{sec:fisher_deri}, while $\delta = 2$ is where the sign of $(W_1^{0,2})_{\|}-(W_1^{0,2})_{\perp}$ changes at the fiducial cosmology, it is also where the dominated RSD effect transits from the Kaiser effect to the FoG effect.

Due to the anisotropies introduced by RSD, distinct information is captured by $(W_i^{0,2})_{\perp}$ and $(W_i^{0,2})_{\|}$, this can be seen for both high- and low-density thresholds in figure~\ref{fig:perp_plus_para}. The perpendicular and parallel elements of MTs show varied sensitivity to the same parameter and different parameter degeneracy for the same parameter pair. Tighter constraints can thus be obtained by the combination of $(W_i^{0,2})_{\perp}$ and $(W_i^{0,2})_{\|}$. We note the perpendicular MTs always give tighter constraints on all parameters than the parallel ones for the low-density thresholds, this is mainly due to smaller uncertainty in $(W^{0,2}_{1,2})_{\perp}$. As has been explained in Section~\ref{sec:3dmts}, it is the average of the first two statistically independent elements of the MTs: $(W^{0,2}_{1,2})_{\perp} = ((W^{0,2}_{1,2})_{1,1}+(W^{0,2}_{1,2})_{2,2})/2$. However, smaller statistical noise doesn't always mean smaller marginalized errors on cosmological parameters. Even with larger variance, $(W^{0,2}_{1,2})_{\|}$ beats $(W^{0,2}_{1,2})_{\perp}$ at the constraint on $\sigma_8$ for high-density thresholds where the FoG effect dominates. 

On the other hand, more distinct and complementary information is embedded in the MTs of low and high-density thresholds. For each of $(W_i^{0,2})_{\perp}$, $(W_i^{0,2})_{\|}$, and their combination, nearly all constraints obtained from statistics of the low thresholds are stronger than those from high-density thresholds. The constraint of $(W_i^{0,2})_{\perp}$ on $|f_{R_0}|^{\lg 2}$ is the only exception, for which the constraints from the low and high thresholds are comparable to each other. It is expected that statistics of low-density regions are sensitive to $f_{R_0}$ due to the smaller screening effect in low-density regions \cite{2018PhRvD..97b3535V,10.1093/mnras/stab1112,Shao_2019}. The comparable constraining power on the f(R) parameter of $(W_i^{0,2})_{\perp}$ for low and high thresholds may indicate rich information about $f_{R_0}$ can also be found from intermediate- and high-density regions. The constraint on $M_{\nu}$ is consistent with \cite{2021ApJ...919...24B}, where they found void size function has stronger constraining power on $M_{\nu}$ than the halo mass function. This can be understood because the effect of massive neutrinos on the LSS is strongest in low-density regions \cite{Massara_2015,Kreisch_2019}. 

To summarize, complementary information is found in the perpendicular and parallel elements of the MTs as well as in the MTs for low- and high-density thresholds. The combination of them will provide the MFs with anisotropic information, and help tighten constraints on the f(R) parameter and other parameters.

\subsection{Information in real- and redshift-space}
\label{sec:iso_aniso}

\begin{figure}[tbp]
	\centering 
	\includegraphics[width=1.0\textwidth]{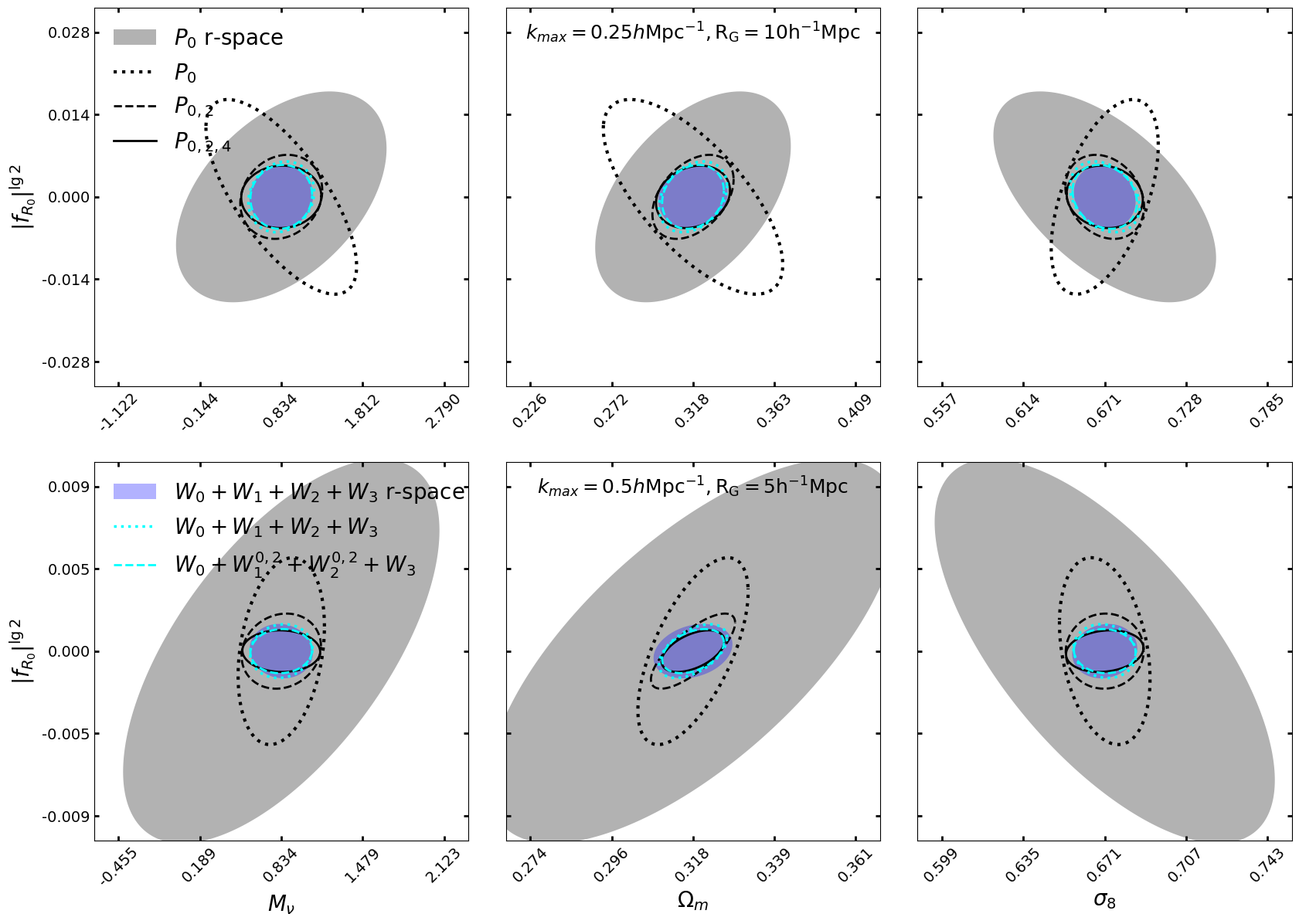}
	\caption{\label{fig:iso_plus_aniso} Parameter constraint contours from the isotropic statistics and the combination of isotropic and anisotropic statistics on large scales ($k_{\rm{max}}=0.25~h\rm{Mpc}^{-1},\ R_G=10~h^{-1}\rm{Mpc}$, the first row) and small scales ($k_{\rm{max}}=0.5~h\rm{Mpc}^{-1},\ R_G=5~h^{-1}\rm{Mpc}$, the second row). Results for $P_0$ and $W_0+W_1+W_2+W_3$ of real space field are shown with grey and blue filled ellipses, while results for statistics in redshift space are plotted with lines: dotted, dashed, and solid black lines for $P_0$, $P_{0,2}$, and $P_{0,2,4}$, dotted and dashed cyan lines for $W_0+W_1+W_2+W_3$ and $W_0+W^{0,2}_1+W^{0,2}_2+W_3$. We note that the results present in this figure is performed for the four parameters $\{|f_{R_0}|^{\lg 2},M_{\nu},\Omega_m,\sigma_8\}$ and the value of $h$ is fixed (see text for more information).}
\end{figure}

For cosmological models with different values of neutrino masses and modified gravity parameters, those with a comparable matter power spectrum at a given time can have different growth rates \cite{2019A&A...629A..46H,Wright_2019}. Motivated by this finding, we will compare the sensitivity of statistics in real space with that in redshift space and investigate how the constraining power on the f(R) parameter, neutrino mass, and other parameters is strengthened when anisotropic information is included. We note that the Fisher forecast present in figure~\ref{fig:iso_plus_aniso} and discussed in this section is performed for the four parameters $\{|f_{R_0}|^{\lg 2},M_{\nu},\Omega_m,\sigma_8\}$, the value of $h$ is fixed because the strong degeneracy between $h$ and $\Omega_m$ in the real-space MFs makes us fail to get a very converged Fisher forecast for the full five parameters. The relation between parameter degeneracies in the statistic and the convergence of its Fisher forecast will be discussed in Appendix~\ref{sec:conver}.

As can be seen in figure~\ref{fig:iso_plus_aniso}, strong degeneracies exist in the real space power spectrum monopole ($P_0$ r-space) between $|f_{R_0}|^{\lg 2}$ and $\{M_{\nu},\Omega_m,\sigma_8\}$ for both large (weakly non-linear) and small (non-linear) scales. These degeneracies are only slightly weakened on large scales but significantly reduced on small scales in the redshift-space power spectrum monopole ($P_0$). Therefore, we observe comparable constraints from $P_0$ in real- and redshift-space on large scales but much tighter constraints from $P_0$ in redshift space than real space, on small scales.

There is a different story for the MFs. Similar parameter degeneracies and constraints are obtained from the MFs in real- and redshift-space for both large and small scales. However, this is only true when the Fisher forecast is performed for $\{|f_{R_0}|^{\lg 2},M_{\nu},\Omega_m,\sigma_8\}$. Extra information is indeed captured by the MFs in redshift-space so that the degeneracy between $\Omega_m$ and $h$ is broken, that is why we can obtain converged Fisher matrix forecasts for the full five parameters when we focus on statistics in redshift space.

When anisotropic information contained in $P_2$ and $P_4$ is added, we can see the parameter degeneracies are gradually reduced, and parameter constraints are improved. The anisotropic information in $P_2$ is important for all parameters on large scales but only important for the f(R) parameter and $\Omega_m$ on small scales. Anisotropies captured by $P_4$ are only helpful for $|f_{R_0}|^{\lg 2}$ and $\Omega_m$ on both large and small scales.

The anisotropic information captured by the MTs can only slightly help improve sensitivity to $|f_{R_0}|^{\lg 2}$ and $\Omega_m$ on both large and small scales. There are three possible explanations for this: first, the MFs already contain considerable non-Gaussian information related to the growth rate so that only a small part of information leaks into anisotropic information; second, the MTs used in this work only capture a small part of anisotropies, we need other independent rank-2 MTs or even higher rank MTs to extract more anisotropies \cite{2022JSMTE2022d3301K}; third, a large part of anisotropies in the redshift space are smeared out by the Gaussian window function, this can be directly seen in figure~\ref{fig:Contours}, where the RSD effect is barely perceptible by human eyes.

\subsection{Comparison and combination of the power spectrum and MTs}
\label{sec:pkvsmt}

\begin{figure}[tbp]
	\centering 
	\includegraphics[width=1.0\textwidth]{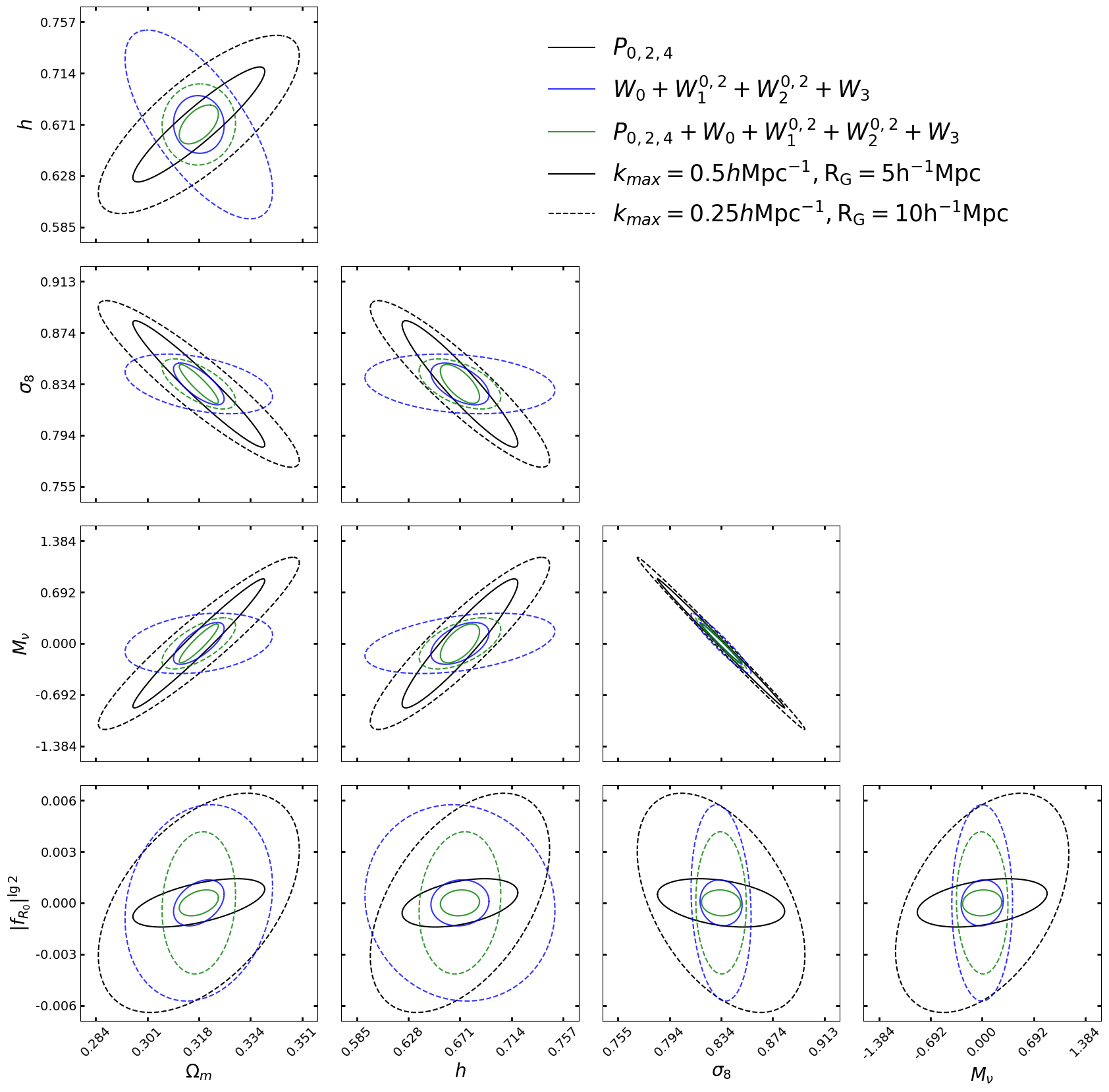}
	\caption{\label{fig:individual} Parameter constraint contours from the power spectrum $P_{0,2,4}$ (black), the MTs $W_0+W^{0,2}_1+W^{0,2}_2+W_3$ (blue), and the combination of the power spectrum and MTs $P_{0,2,4}+W_0+W^{0,2}_1+W^{0,2}_2+W_3$ (green). Results for small scales ($k_{\rm max}=0.5~h\rm{Mpc}^{-1},\ R_G=5~h^{-1}\rm{Mpc}$) and larger scales ($k_{\rm max}=0.25~h\rm{Mpc}^{-1},\ R_G=10~h^{-1}\rm{Mpc}$) are plotted in solid and dashed lines, respectively.}
\end{figure}

\begin{center}
	\small
	\begin{table}[tbp]
		\scalebox{0.758}[1]{
			\begin{tabular}{|c|c|c|c|c|c|c|}
				\hline
                            Statistic & $P_{0,2,4}$ & $W_0+W^{0,2}_{1,2}+W_3$ & $P_{0,2,4}+W_0+W^{0,2}_{1,2}+W_3$ & $P_{0,2,4}$ & $W_0+W^{0,2}_{1,2}+W_3$ & $P_{0,2,4}+W_0+W^{0,2}_{1,2}+W_3$ \\
                             \hline Scale & \multicolumn{3}{|c|}{$k_{\rm max}=0.5~h\rm{Mpc}^{-1},\ R_G=5~h^{-1}\rm{Mpc}$} & \multicolumn{3}{|c|}{$k_{\rm max}=0.25~h\rm{Mpc}^{-1},\ R_G=10h^{-1}\rm{Mpc}$} \\
                            \hline$\Omega_m$ & 0.014 & 0.005 & 0.004 & 0.022 & 0.016 & 0.008 \\
                            $h$ & 0.032 & 0.016 & 0.011 & 0.049 & 0.053 & 0.022 \\
                            $\sigma_8$ & 0.032 & 0.011 & 0.010 & 0.043 & 0.016 & 0.013 \\
                            $M_\nu$ & 0.572 & 0.185 & 0.171 & 0.766 & 0.276 & 0.223 \\
                            $|f_{R_0}|^{\lg 2}$ & 0.0009 & 0.0008 & 0.0005 & 0.0040 & 0.0036 & 0.0026 \\
                            \hline
			\end{tabular}
		}
		\caption{\label{tab:Pk_plus_other} Marginalized parameter constraints for statistics on small scales (the first three columns, $k_{\rm max}=0.5~h\rm{Mpc}^{-1},\ R_G=5~h^{-1}\rm{Mpc}$) and large scales (the last three columns, $k_{\rm max}=0.25~h\rm{Mpc}^{-1},\ R_G=10~h^{-1}\rm{Mpc}$), from the power spectrum $P_{0,2,4}$ (first and forth columns), the MTs $W_0+W^{0,2}_{1,2}+W_3$ (second and fifth columns), and the combination of power spectrum and MTs $P_{0,2,4}+W_0+W^{0,2}_{1,2}+W_3$ (third and last columns).} 
	\end{table}
\end{center}

\begin{center}
	\small
	\begin{table}[tbp]
		\scalebox{0.675}[1]{
			\begin{tabular}{|c|c|c|c|c|}
				\hline
                            Statistic & $P_{0,2,4}/(W_0+W^{0,2}_{1,2}+W_3)$ & $P_{0,2,4}/(P_{0,2,4}+W_0+W^{0,2}_{1,2}+W_3)$ & $P_{0,2,4}/(W_0+W^{0,2}_{1,2}+W_3)$ & $P_{0,2,4}/(P_{0,2,4}+W_0+W^{0,2}_{1,2}+W_3)$ \\
                             \hline Scale & \multicolumn{2}{|c|}{$k_{\rm max}=0.5~h\rm{Mpc}^{-1},\ R_G=5~h^{-1}\rm{Mpc}$} & \multicolumn{2}{|c|}{$k_{\rm max}=0.25~h\rm{Mpc}^{-1},\ R_G=10~h^{-1}\rm{Mpc}$} \\
                            \hline$\Omega_m$ & 2.6 & 3.4 & 1.4 & 2.8 \\
                            $h$ & 2.0 & 3.0 & 0.9 & 2.2 \\
                            $\sigma_8$ & 3.0 & 3.3 & 2.7 & 3.4 \\
                            $M_\nu$ & 3.1 & 3.3 & 2.8 & 3.4 \\
                            $|f_{R_0}|^{\lg 2}$ & 1.0 &1.9 & 1.1 & 1.5 \\
                            \hline
			\end{tabular}
		}
		\caption{\label{tab:Ratio_pk_to_other} Marginalized parameter constraint improvements for the power spectrum on small scales (the first two columns, $k_{\rm max}=0.5~h\rm{Mpc}^{-1},\ R_G=5~h^{-1}\rm{Mpc}$) and large scales (the last two columns, $k_{\rm max}=0.25~h\rm{Mpc}^{-1},\ R_G=10~h^{-1}\rm{Mpc}$), from the  MTs $W_0+W^{0,2}_{1,2}+W_3$ (first and third columns), and the combination of power spectrum and MTs $P_{0,2,4}+W_0+W^{0,2}_{1,2}+W_3$ (second and last columns).} 
	\end{table}
\end{center}

We have shown the improvement from MTs is not effective for MFs. In this section, we will focus on how the non-Gaussian information from the MTs (``MTs'' will refer to $W_0+W_1^{0,2}+W_2^{0,2}+W_3$ hereafter in this section) helps break the degeneracies between $|f_{R_0}|^{\lg 2}$ and $M_{\nu}$ and between other parameter pairs on both large and small scales.  The marginalized $68\%$ Fisher matrix confidence contours from the power spectrum multipoles ($P_{0,2,4}$), MTs ($W_0+W_1^{0,2}+W_2^{0,2}+W_3$), combination of power spectrum and MTs ($P_{0,2,4}+W_0+W_1^{0,2}+W_2^{0,2}+W_3$) on large ($k_{\rm max}=0.25~h\rm{Mpc}^{-1},\ R_G=10~h^{-1}\rm{Mpc}$) and small ($k_{\rm max}=0.5~h\rm{Mpc}^{-1},\ R_G=5~h^{-1}\rm{Mpc}$) scales are shown in figure~\ref{fig:individual}. The corresponding parameter constraints from these statistics on the two scales are also listed in table~\ref{tab:Pk_plus_other}.  For a better comparison, the ratio of constraints from the power spectrum to those from the MTs as well as to those from the combination $P_{0,2,4}+W_0+W_1^{0,2}+W_2^{0,2}+W_3$ is present in table~\ref{tab:Ratio_pk_to_other}, also for both the small and large scales.

Before comparing the constraining power of the power spectrum and MTs, we shall clarify that there is no exact correspondence between the smoothing scale used for the MTs and the $k_{\rm max}$ choice for the power spectrum. In practice, the smallest smoothing scale $R_G=5~h^{-1}\rm{Mpc}$ is the scale where we believe the algorithm used to measure the MTs is accurate enough (see Section~\ref{sec:3dmts} for a more detailed explanation). We also note that whether the modes $k>0.5~h\rm{Mpc}^{-1}$ are included or not only make a $0\%$, $1\%$, $2\%$, and $3\%$ difference in the amplitude of the $W_0$, $W_1$, $W_2$, and $W_3$ for $R_G=5~h^{-1}\rm{Mpc}$. This indicates the information in the MTs may be indeed dominated by modes $k\leq 0.5~h\rm{Mpc}^{-1}$. A fairer comparison between constraints from the power spectrum and the MTs could be made if we also implemented a sharp k-filter on top of the Gaussian smoothing following \cite{2024arXiv240718647V}. We have done a test with the MFs in Appendix~\ref{sec:k-cut}, where we find the sharp k-filter only lead to a slight variation in our results. 

On large scales, the constraints from $P_{0,2,4}$ on $\Omega_m$, $h$, and $|f_{R_0}|^{\lg 2}$ are comparable with those from the MTs, and the MTs have stronger constraining power on $M_{\nu}$ and $\sigma_8$. The overall smaller degeneracies among all parameters (it is hard to see whether the degeneracy between $M_{\nu}$ and $\sigma_8$ become weaker or not for the MTs in figure~\ref{fig:individual}, but we find the correlation coefficient between $M_{\nu}$ and $\sigma_8$ is -0.997 for $P_{0,2,4}$ and -0.958 for the MTs) in the MTs significantly help tighten parameter constraints from $P_{0,2,4}$ by a factor of 2.8, 2.2, 3.4, 3.4, and 1.5 for $\Omega_m$, $h$, $\sigma_8$, $M_{\nu}$, $|f_{R_0}|^{\lg 2}$, when combining $P_{0,2,4}$ and the MTs.

On small scales, the velocity information captured by $P_{0,2,4}$ breaks the degeneracy between the f(R) parameter and other parameters. A comparable constraint on $|f_{R_0}|^{\lg 2}$ with that from the MTs can thus be achieved with the power spectrum alone. However, strong degeneracies still exist in the power spectrum among $\Omega_m$, $h$, $\sigma_8$, and $M_{\nu}$. The degeneracies in the MTs among all parameters are smaller than those in $P_{0,2,4}$, which is the same as what is found on large scales. 
Non-Gaussian information captured by the MTs is more sensitive to the four parameters apart from $|f_{R_0}|^{\lg 2}$, and 2.6, 2.0, 3.0, and 3.1 times tighter constraints are obtained from the MTs alone than those from the power spectrum.  
The high sensitivity of the MTs to the four parameters, together with the different degeneracies among parameters in the MTs from those in $P_{0,2,4}$, are why a factor of 3.4, 3.0, 3.3, 3.3, and 1.9 of improvement in the power spectrum constraints on $\Omega_m$, $h$, $\sigma_8$, $M_{\nu}$, and $|f_{R_0}|^{\lg 2}$ is achieved by the combination of $P_{0,2,4}$ and the MTs.

Focusing on the f(R) parameter and neutrino mass, first on large scales, the MTs help break the $|f_{R_0}|^{\lg 2}-M_{\nu}$ degeneracy existing in the $P_{0,2,4}$ but its constraint on $|f_{R_0}|^{\lg 2}$ is moderately improved by a factor of 1.5 by the combination of the power spectrum and MTs. This improvement is smaller than that on small scales, where the constraint of $P_{0,2,4}$ on $|f_{R_0}|^{\lg 2}$ is improved by a factor of 1.9 because more non-Gaussian information is contained in the MTs on small scales than on large scales. We have shown that the distinct imprints left on the LSS by f(R) gravity in \cite{2017PhRvL.118r1301F} and by massive neutrinos in \cite{2022JCAP...07..045L} can be captured by the MFs on small scales. By the measurement of MFs at multiple density thresholds, the effect of f(R) gravity and massive neutrinos on voids, filaments, and high-density regions is detected by the MFs, and thus by the MTs.

\section{Discussions}
\label{sec:discuss}

Also based on the Quijote and Quijote-MG simulations, \cite{2024arXiv240718647V} conducted Fisher forecasts using the WST coefficients and the power spectrum in real space at $z=0$. The Fisher analysis is performed for $\{\Omega_m,\ \Omega_b,\ h,\ n_s,\ \sigma_8,\ M_{\nu},\ |f_{R_0}|^{\lg 2}\}$, therefore the comparison between results present in this work and that work may not be fair because the degeneracies between $\Omega_m,\ h,\ \sigma_8,\ M_{\nu},\ |f_{R_0}|^{\lg 2}$ and $\Omega_b,\ n_s$ are not included in this work. We find tighter constraints on $|f_{R_0}|^{\lg 2}$ from the power spectrum multipoles and MTs in redshift space compared with that from the power spectrum monopole and WST coefficients. This is expected because the velocity information captured by the power spectrum multipoles and MTs helps increase their sensitivity to the f(R) parameter. In addition, the degeneracies between $f_{R_0}$ and $\Omega_b,\ n_s$
may also weaken the constraints on the f(R) parameter obtained from the WST coefficients.

The idea of using kinematic information to break the degeneracy between $M_{\nu}$ and $f_{R_0}$ was first proposed in \cite{2019A&A...629A..46H}, they found the kinematic information related to both the growth rate on large scales and the virial velocities inside of collapsed structures are well suited to constrain deviations from general relativity without being affected by the effect of massive neutrinos. This is consistent with our findings in Section~\ref{sec:info_MTs}, we find both the anisotropic information captured by the MTs of low-density regions, where the Kaiser effect dominates, and that captured by the MTs of high-density regions, where the FoG effect dominates, are sensitive to the f(R) parameter in the presence of massive neutrinos. Following this idea, \cite{Wright_2019,10.1093/mnras/stz1850} investigated the possibility of breaking the degeneracy between massive neutrinos and f(R) gravity model using RSD with the monopole and quadrupole of 2-point statistics in redshift space. We have discussed this topic in Section~\ref{sec:iso_aniso} and found the degeneracy between $M_{\nu}$ and $f_{R_0}$ can be reduced even with the monopole of power spectrum alone in redshift space, the degeneracy can be broken further when the quadrupole is included, and there still exists extra information about $f_{R_0}$ in the hexadecapole. In addition, the anisotropies captured by the MTs can improve the constraining power of the MFs on the f(R) parameter.

Alternative statistics apart from the power spectrum and two-point correlation functions of particles, halos, or galaxies are also explored for 3D LSS \cite{2018PhRvD..97b3535V,10.1093/mnras/stab1112,2020ApJ...904...93R,Lee_2022,Lee_2023,2023A&A...674A.185M}. However, a direct comparison between their results and ours is difficult because none of them has derived a constraint on $M_{\nu}$ and $f_{R_0}$ at the same time. Section~\ref{sec:pkvsmt} has quantified the improvement from non-Gaussian information captured by the MTs for both non-linear and quasi-linear scales. We find the non-Gaussian information is important to constrain $M_{\nu}$ and $f_{R_0}$ at the same time on these scales.


\section{Conclusions}
\label{conclusions}
We are at the beginning of the Stage-IV era of precision cosmology, the current and upcoming cosmological observations of the cosmic large-scale structure of the Universe may enable the detection of massive neutrinos and modifications of the gravity model at the same time. In this work, we investigate the possibility of tightly constraining the mass sum of neutrinos and the f(R) gravity model simultaneously with the structure and velocity information extracted using the power spectrum multipoles and Minkowski tensors. 

As the first application of the tensor Minkowski functionals in cosmology on fully non-linear scales, we first describe and interpret in detail how the effect of redshift-space distortions on low, intermediate, and high-density regions is captured by the 3D MTs; then the influence of $f_{R_0}$, $M_{\nu}$, and other cosmological parameters on the MTs are compared and the non-degenerate imprints of $f_{R_0}$ and $M_{\nu}$ are found on the MTs; finally, to analyze the information embedded in the MTs, we quantify the information content in the perpendicular and parallel element of the MTs for both low and high-density regions with the Fisher information formalism, which helps us understand how the anisotropic information dominated by the Kaiser and FoG effect can put constraints on $f_{R_0}$ and $M_{\nu}$ and other cosmological parameters.

Motivated by previous works, we also study how the velocity information helps break the degeneracy between $f_{R_0}$ and $M_{\nu}$. The velocity information is added into the CDM field by the redshift-space distortion, which is then measured by all of the statistics used in this work. The power spectrum monopole, although as an isotropic statistic, can still detect the effect of RSD and thus extract a part of velocity information. Therefore, we see the $f_{R_0}$ and $M_{\nu}$ degeneracy is already reduced in the redshift space power spectrum monopole. However, the constraining power of the MFs mainly comes from the captured non-Gaussianities, no significant difference in the sensitivity to $f_{R_0}$ and $M_{\nu}$ is seen between real- and redshift-space. With the quadrupole and hexadecapole of the power spectrum and the MTs, more velocity information is extracted and stronger constraints are obtained, especially for the f(R) parameter. However, we find that the MTs provide only limited supplementary information to the MFs. On the other hand, measuring the MTs requires generating complex isodensity contours and performing time-consuming computations over them, which incurs over 700 times the computational cost of measuring the MFs using grid-based algorithms. Therefore, there is a need to develop a more efficient algorithm for measuring the MTs and to explore higher-rank MTs in order to capture more anisotropic information in the field.

To extend our previous works, we do a comparison between constraints from the power spectrum and the MTs instead of the MFs alone and investigate how the combination of the power spectrum and the MTs helps improve the constraints from $P_{0,2,4}$. We find that the non-Gaussian information indeed is helpful to break the $f_{R_0}$ and $M_{\nu}$ degeneracy, which is consistent with other works based on other non-Gaussian statistics. The MTs, similar to what has been found for the MFs \cite{2021arXiv210803851J,2022JCAP...07..045L,2023JCAP...09..037L}, can provide complementary information for the power spectrum and tighten the constraints significantly for all of the parameters considered in this work on both quasi- and non-linear scales.

Tighter constraints can be obtained if we extend our analysis to smaller scales, as indicated by Figure 7 of \cite{2024arXiv240718647V}. However, we choose only to include k-modes up to $k_{\rm{max}}=0.5~h\rm{Mpc}^{-1}$, considering the finite resolution of simulations used in this work. For the MTs, the smoothing scale should be at least larger than the average distance between particles \cite{1994ApJ...420..525V,1989ApJ...340..625G,hikage2003minkowski,2005ApJ...633....1P}, which is about $2h^{-1}\rm{Mpc}$ for the simulations used here. On the other hand, the smoothing scale $R_G=5h^{-1}\rm{Mpc}$ is the smallest scale we can use to get rid of the discretization error of iso-density contour generation algorithm used in this work \cite{2018ApJ...863..200A}. A finer density field is required if we want to extend the analysis to scales smaller than $R_G=5h^{-1}\rm{Mpc}$ and extend the analysis to higher-rank Minkowski tensors capable of extracting more anisotropic information. However, the computation time will exceed what we can afford for such an analysis. Better results can also be calculated by combining the MTs with different smoothing scales, we don't discuss it in this work because it is already studied in our previous work \cite{2022JCAP...07..045L,2023JCAP...09..037L}. A more interesting extension of this work will be to combine statistics at different redshift because it has been found in previous works \cite{10.1093/mnras/stab1112,10.1093/mnras/stz1850} that the degeneracy between massive neutrinos and the f(R) gravity may be more easily disentangled at higher redshifts.

The reliability of the Fisher forecast depends on both the Gaussianity of the likelihood of the statistics and the convergence of results w.r.t. the noise in the derivatives and covariance matrix as well as the inherent error in the numerical differentiation methods. We have discussed these topics in Appendices~\ref{sec:Gaussian_test}, \ref{sec:conver}, \ref{sec:test_deri} and showed our analysis presented in this work has passed all of these tests. A more conservative result which has the best convergence is presented and discussed in Appendix~\ref{sec:conservative_result}. We have also investigated the fairness of comparing the constraining power of the MTs measured for $R_G=5,10\ h^{-1}\rm Mpc$ with the power spectrum of modes up to $k_{\rm max}=0.5,0.25\ h \rm Mpc^{-1}$ in Appendix~\ref{sec:k-cut}.

In reality, the observed galaxy positions are distorted by their peculiar velocities along LoS and the LoS vector varies at different points in space. This differs from the plane-parallel approximation used in this work when applying the RSD in the simulations, where a common LoS vector is shared at each point in the field. However, the effect of RSD on the MTs beyond the plane-parallel approximation is discussed only on linear and quasi-linear scales in \cite{2022arXiv220810164A}. Besides the RSD effect, the Alcock-Paczynski (AP) distortions, the complicated geometry, and other systematics of galaxy surveys can introduce extra anisotropies into the field. In addition, the model of the halo-galaxy connection and the emulation of their dependence on both cosmological and halo-galaxy connection parameters should also be taken into consideration in order to apply the MTs to galaxy catalogs from spectroscopic surveys. We leave the investigation for these topics to future work.

\acknowledgments
WL appreciated the insightful discussions with Aoxiang Jiang. The work of WL and WF is supported by the National Natural Science Foundation of China Grants No. 12173036 and 11773024, by the National Key R$\&$D Program of China Grant No. 2021YFC2203100 and No. 2022YFF0503404, by the China Manned Space Project “CSST Cluster Precise Mapping and Cluster Cosmology” and Grant No. CMS-CSST-2021-B01, by the Fundamental Research Funds for Central Universities Grants No. WK3440000004 and WK3440000005, by Cyrus Chun Ying Tang Foundations, and by the 111 Project for "Observational and Theoretical Research on Dark Matter and Dark Energy" (B23042). LW is supported by National Key R$\&$D Program of China No. 2023YFA1607903, National Natural Science Foundation of China (grant nos. 12033006 $\&$ 12192221), and the Cyrus Chun Ying Tang Foundations. GV acknowledges the support of the Eric and Wendy Schmidt AI in Science Postdoctoral Fellowship at the University of Chicago, a Schmidt Sciences, LLC program.

\appendix

\section{Gaussianity test}
\label{sec:Gaussian_test}
\begin{figure}[tbp]
	\centering
	\includegraphics[width=1.0\textwidth]{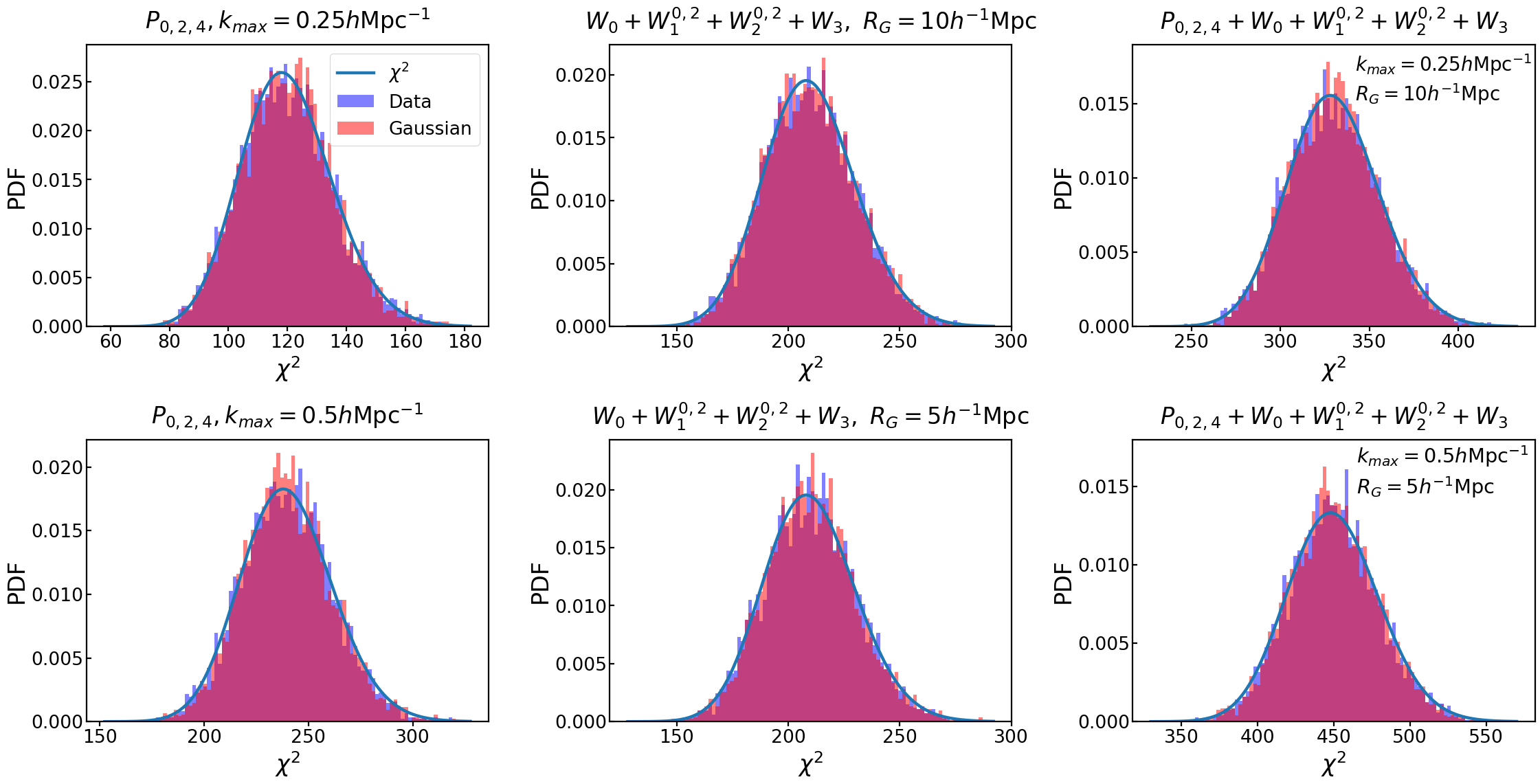}
	\caption{\label{fig:gaussian_test} A qualitative assessment of the Gaussianity of the likelihood for the power spectrum multipoles ($P_{0,2,4}$, first column), the MTs ($W_0+W_1^{0,2}+W_2^{0,2}+W_3$, second column), and their combination ($P_{0,2,4}+W_0+W_1^{0,2}+W_2^{0,2}+W_3$, third column) for large ($k_{\rm{max}}=0.25~h\rm{Mpc}^{-1}$, $R_G=10~h^{-1}\rm{Mpc}$, first row) and small ($k_{\rm{max}}=0.5~h\rm{Mpc}^{-1}$, $R_G=5~h^{-1}\rm{Mpc}$, second row) scales. The histograms of the $\chi^2$ values measured from the Quijote simulations are drawn in blue, while those of the $\chi^2$ values measured from a multivariate Gaussian distribution with the same mean and covariance as the Quijote simulations are shown in red. The solid lines show theoretical $\chi^2$ distributions with degrees of freedom equal to the length of observables.} 
\end{figure}

It was reported in \cite{2022arXiv220405435P} that non-Gaussianities in the likelihood of statistics might lead to artificially tight bounds on the cosmological parameters using the Fisher matrix formalism. To access the Gaussianity of the likelihood of the statistics used in this work, we follow the analysis performed in \cite{10.1093/mnras/stab2384,2022arXiv220904310P} and check that the likelihood of these statistics can be approximated by the multivariate Gaussian. Since we have 5000 fiducial simulations, we can obtain 5000 $\chi^2$ values for each of the statistics by 
\begin{equation}
\chi^2_i=(\boldsymbol{d_i}-\boldsymbol{\mu})^T C^{-1} (\boldsymbol{d_i}-\boldsymbol{\mu}),
\end{equation}
where $\boldsymbol{d_i}$ is the data vector of the summary statistics for the $i$-th simulation, $\boldsymbol{\mu}$ and $C$ is the mean and the covariance matrix of the data vector estimated from the 5000 simulations. 

If the assumption of Gaussian likelihood holds, the $\chi^2$ values are expected to follow a $\chi^2$ distribution with degrees of freedom equal to the length of the data vector. In figure~\ref{fig:gaussian_test}, we plot the histogram (in blue) of the $\chi^2$ values measured from the Quijote simulations and compare it with the theoretical $\chi^2$ distribution curve for each of the summary statistics and their combination. Due to the existence of fluctuations in the histogram, the curve may not exactly agree with the histogram, even for a sample strictly following the $\chi^2$ distribution. To visualize the amplitude of the fluctuations around the theoretical $\chi^2$ distribution curve and use it as a ruler to help assess the Gaussianity of the data vectors, we create 5000 multivariate Gaussian distributed data vectors with the same mean and covariance matrix as those estimated from the fiducial simulations. We then obtain 5000 $\chi^2$ values for the multivariate Gaussian distributed data vectors and also plot a histogram (in red) for them in figure~\ref{fig:gaussian_test}. As seen in this figure, the histogram of $\chi^2$ values for all statistics is very close to that for the Gaussian distributed data vectors and agrees well with the theoretical $\chi^2$ distribution curve. This indicates that the likelihood of the power spectrum, MTs, and their combination on both small and large scales can be well modeled as Gaussian.

\section{Convergence w.r.t. noise in derivatives and covariance matrix}
\label{sec:conver}

\begin{figure}[tbp]
	\centering
	\includegraphics[width=1.0\textwidth]{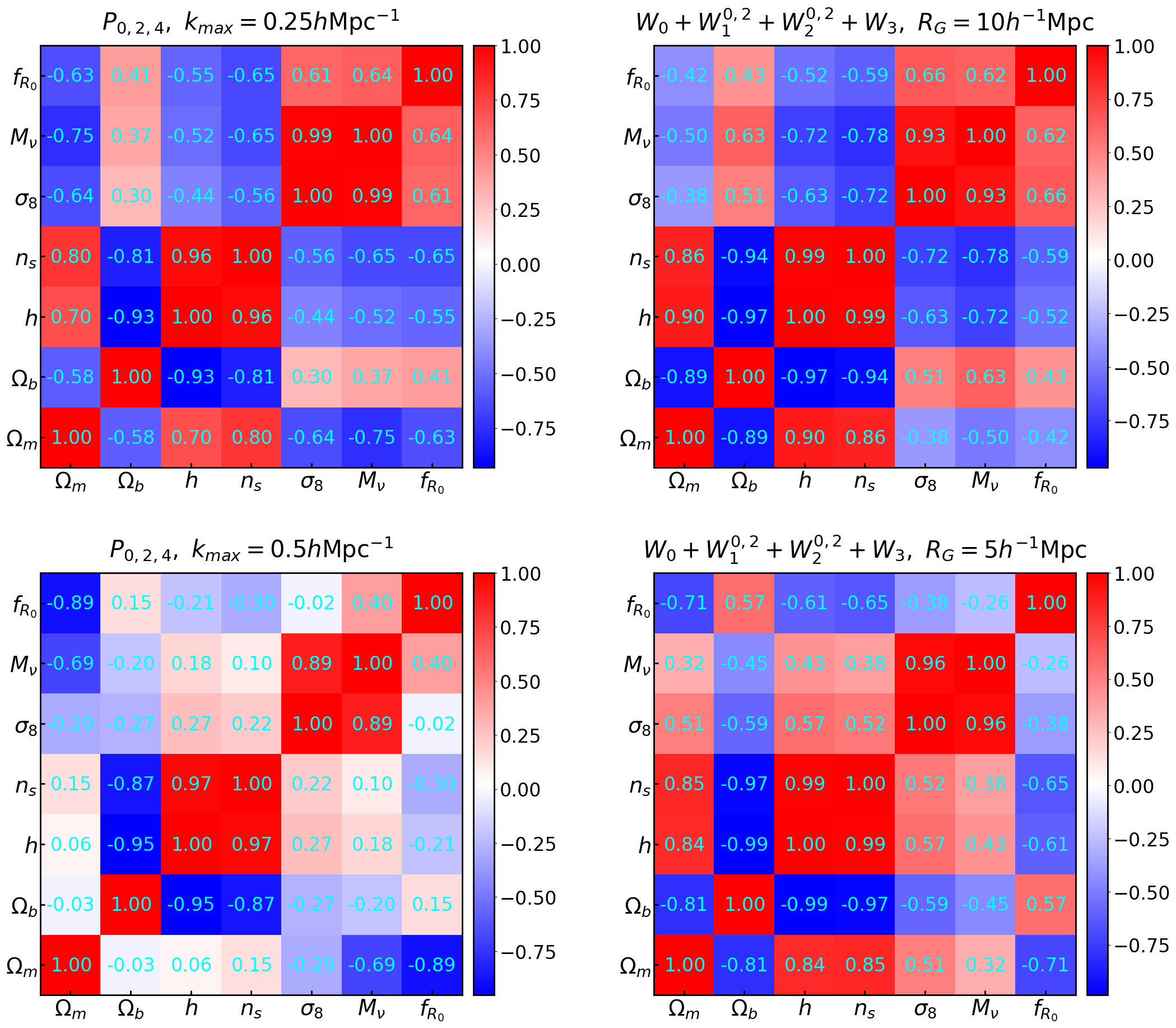}
	\caption{\label{fig:Fisher_correlation} The ``Fisher correlation matrix'' calculated using $P_{0,2,4}$ with $k_{\rm{max}}=0.25~h\rm{Mpc}^{-1}$ (top left panel) and $k_{\rm{max}}=0.5~h\rm{Mpc}^{-1}$ (bottom left panel) as well as using $W_0+W_1^{0,2}+W_2^{0,2}+W_3$ with $R_{G}=10h^{-1}\rm{Mpc}$ (top right panel) and $R_{G}=5h^{-1}\rm{Mpc}$ (bottom right panel) for $\Omega_{m}$, $\Omega_{b}$, $h$, $n_s$, $\sigma_{8}$, $M_{\nu}$ and $f_{R_0}$.} 
\end{figure}

\begin{figure}[tbp]
	\centering
	\includegraphics[width=1.0\textwidth]{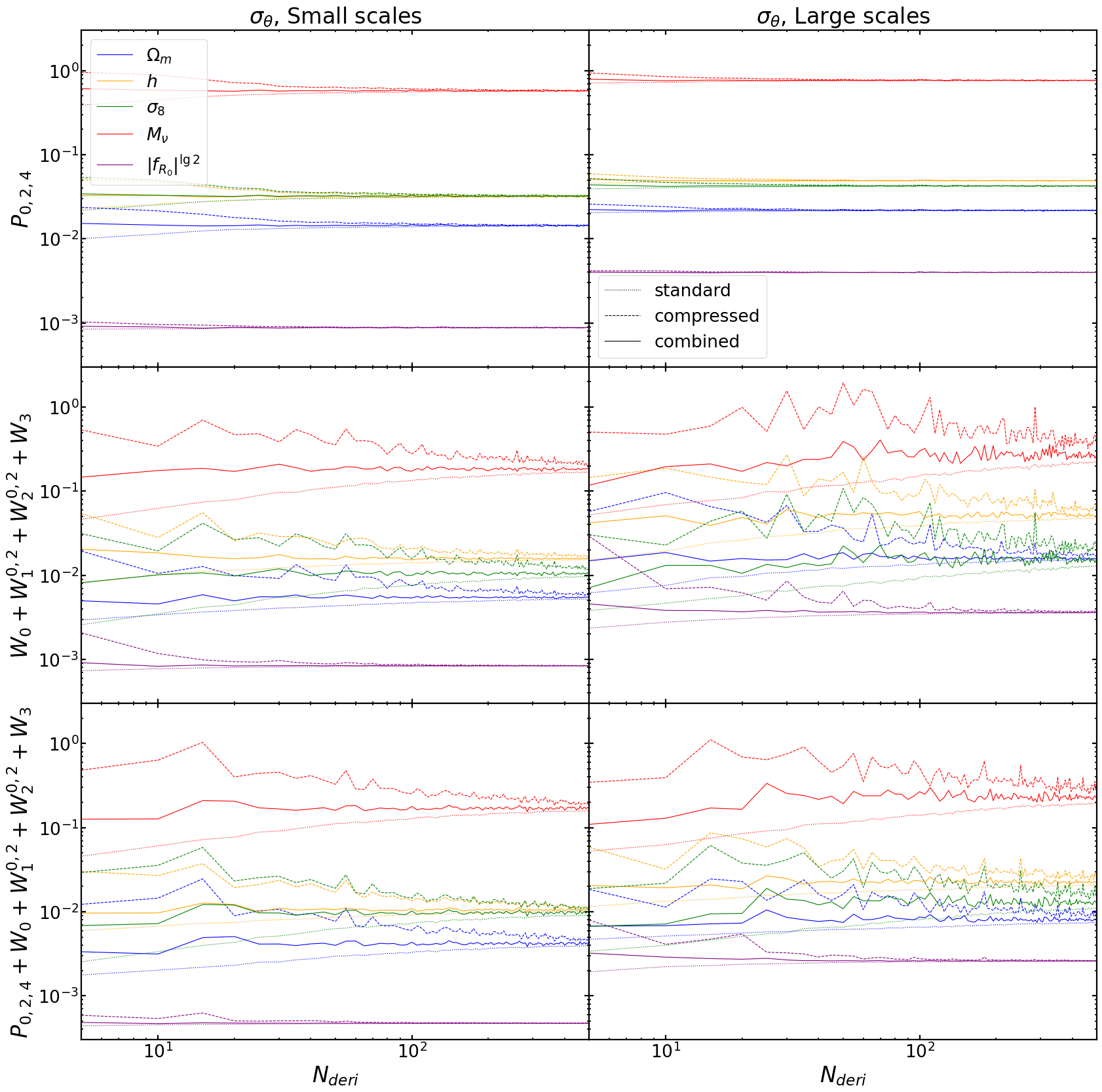}
	\caption{\label{fig:Convergence_test_deri_5p} Convergence of the marginalized errors from the power spectrum multipoles ($P_{0,2,4}$, first row), the MTs ($W_0+W^{0,2}_{1}+W^{0,2}_{1}+W_3$, second row), and their combination ($P_{0,2,4}+W_0+W^{0,2}_{1}+W^{0,2}_{1}+W_3$, third row) on $\Omega_{m}$, $h$, $\sigma_{8}$, $M_{\nu}$ and $f_{R_0}$ (blue, orange, green, red, purple) for both small ($k_{\rm{max}}=0.5~h\rm{Mpc}^{-1},\ R_G=5~h^{-1}\rm{Mpc}$, left column) and large ($k_{\rm{max}}=0.25~h\rm{Mpc}^{-1},\ R_G=10~h^{-1}\rm{Mpc}$, right column) scales.  The Fisher forecasts are obtained with derivatives estimated from $N_{deri}$ simulations using the standard, compressed, and combined estimator (dotted, dashed, and solid lines). These forecasts are all based on covariance matrices estimated with $N_{cov}=5000$ simulations.} 
\end{figure}

\begin{figure}[tbp]
	\centering
	\includegraphics[width=1.0\textwidth]{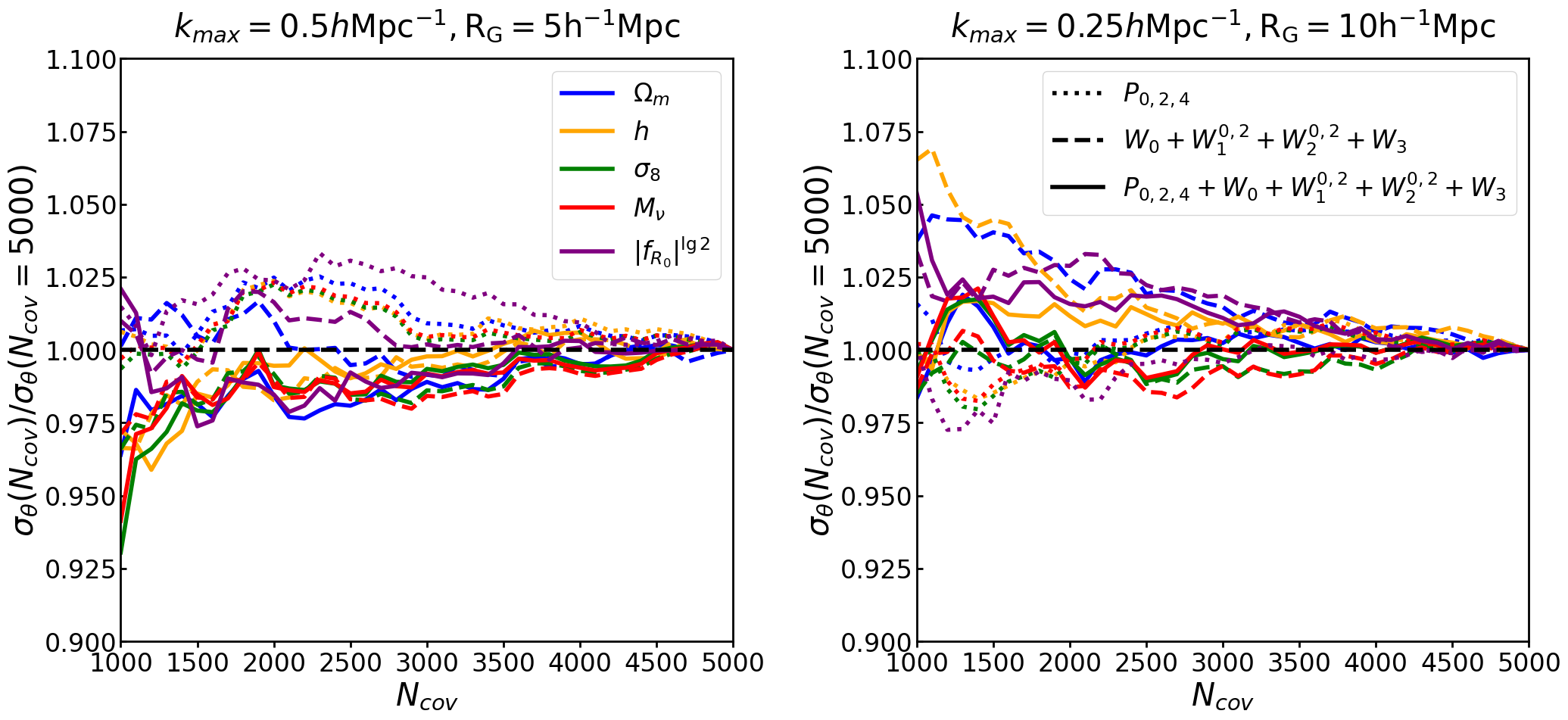}
	\caption{\label{fig:Convergence_test_Ncov} Convergence of the marginalized errors from the power spectrum multipoles ($P_{0,2,4}$, dotted lines), the MTs ($W_0+W^{0,2}_{1}+W^{0,2}_{2}+W_3$, dashed lines), and their combination ($P_{0,2,4}+W_0+W^{0,2}_{1}+W^{0,2}_{2}+W_3$, solid lines) on $\Omega_{m}$, $h$, $\sigma_{8}$, $M_{\nu}$ and $f_{R_0}$ (in blue, orange, green, red, purple), for both small ($k_{\rm{max}}=0.5~h\rm{Mpc}^{-1},\ R_G=5~h^{-1}\rm{Mpc}$, left column) and large ($k_{\rm{max}}=0.25~h\rm{Mpc}^{-1},\ R_G=10~h^{-1}\rm{Mpc}$, right column) scales. $\sigma_{\theta}(N_{cov})/\sigma_{\theta}(N_{cov}=5000)$ is the ratio of Fisher forecasts obtained with covariance matrices estimated from $N_{cov}$ simulations to those obtained with covariance matrices estimated from $N_{cov}=5000$ simulations. These forecasts are based on derivatives estimated with $N_{deri}=500$ simulations. The horizontal dashed lines correspond to $\sigma_{\theta}(N_{cov})/\sigma_{\theta}(N_{cov}=5000) = 1$.} 
\end{figure}

\begin{figure}[tbp]
	\centering
	\includegraphics[width=1.0\textwidth]{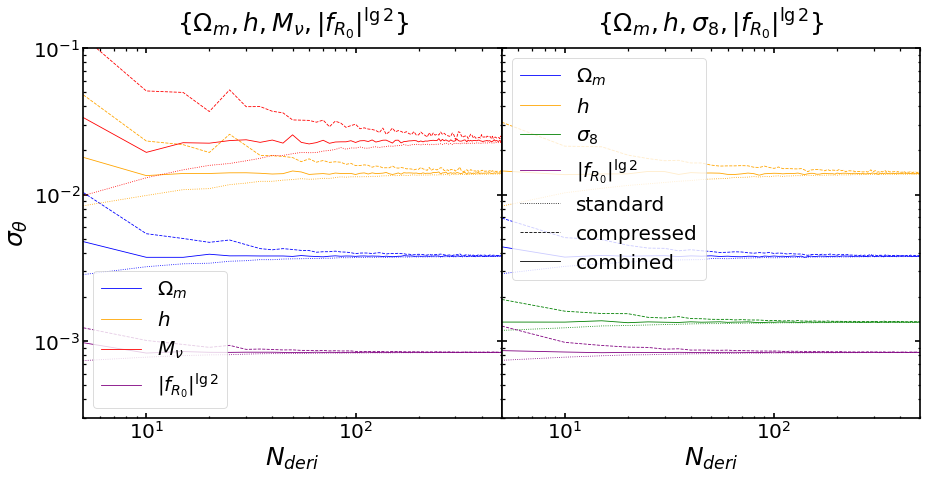}
	\caption{\label{fig:Convergence_deri_4p} Convergence of the marginalized errors from the MTs ($W_0+W^{0,2}_{1}+W^{0,2}_{1}+W_3$) when the Fisher analysis is performed for $\{\Omega_m,\ h,\ M_{\nu},\ f_{R_0}^{\lg 2}\}$ (left column) and $\{\Omega_m,\ h,\ \sigma_8,\ f_{R_0}^{\lg 2}\}$ (right column). The Fisher forecasts are obtained with derivatives estimated from $N_{deri}$ simulations using the standard, compressed, and combined estimator (dotted, dashed, and solid lines). These forecasts are all based on covariance matrices estimated with $N_{cov}=5000$ simulations.} 
\end{figure}

The Fisher matrix forecast, as a computationally efficient and reliable tool, has been widely used in cosmology to estimate the precision with which model parameters can be measured \cite{2023A&A...675A.120E,2010A&A...523A...1J,10.1093/mnras/stx1209,2023ApJ...943..178C}. It can be simply calculated for a set of observables, if both the covariance matrix of these observables and derivatives of each observable w.r.t. parameters are known. However, the noise and bias in both the covariance matrix and derivatives can be propagated into the Fisher matrix, leading to unreliable and biased forecasts \cite{10.1093/mnras/stt270,PhysRevD.88.063537,2023arXiv230508994C,2024arXiv240606067W}.
Here in this appendix, we will explain in which situation the Fisher forecasts are most prone to noise, and convergence tests will be done to show why our forecast is converged and reliable. 

With the Fisher matrix, the marginalized parameter constraints are given by
\begin{equation}
\sigma_{\alpha}=\sqrt{(F^{-1})_{\alpha \alpha}}.
\end{equation}
Let us write the Fisher matrix as the following block matrix structure
\begin{equation}
\left[F_{\alpha \beta}\right]=\left[\begin{array}{ll}F_{\mu \nu} & F_{\mu b} \\ F_{a \nu} & F_{a b}\end{array}\right],
\end{equation}
the determinant of the Fisher matrix is then given by 
\begin{equation}
\det (F_{\alpha \beta})=\det (F_{\mu \nu}) \det( F_{a b} -  F_{a \nu}F^{-1}_{\nu \mu}F_{\mu b}).
\end{equation}
When the matrix block $F_{\mu \nu}$ is calculated for two strongly degenerate parameters like $\sigma_8$ and $M_{\nu}$, $\det (F_{\mu \nu})$ will be very close to zero, and the determinant of the full Fisher matrix $\det (F_{\alpha \beta})$ will also approach zero. Since the parameter covariance matrix is obtained by inverting the Fisher matrix; thus, small uncertainties in the Fisher matrix may result in much larger deviations in the parameter covariance matrix. 
The degeneracies can be detected in advance with the ``Fisher correlation matrix'' 
\begin{equation}
f_{\alpha \beta}=F_{\alpha \beta}/\sqrt{F_{\alpha \alpha}F_{\beta \beta}}.
\end{equation}
If the non-diagonal element of $f_{\alpha \beta}$ is close to -1 or 1, it indicates the degeneracy between the two corresponding parameters may lead to an amplification of error for the inverse of the Fisher matrix. In Figure~\ref{fig:Fisher_correlation}, we show the Fisher correlation matrix for both the power spectrum and MTs on both large and small scales. Strong degeneracies can be seen in all of the four panels among $\Omega_b$, $h$,  and $n_s$, as well as between $\sigma_8$ and $M_{\nu}$. To get a convergent and reliable Fisher forecast, we have to fix two of the three parameters $\Omega_b$, $h$, and $n_s$ and use Eq~\ref{eq:nu_deri} for the estimate of derivatives w.r.t. $M_{\nu}$ instead of the higher order numerical differentiation schemes to reduce the influence of the derivative noise \cite{2024arXiv240606067W}.

The uncertainties of the Fisher matrix stem from the noises existing in both the estimated derivatives and covariance matrix. We will check how the marginalized errors vary when increasing the number of simulations used to estimate the derivatives $N_{deri}$ in figure~\ref{fig:Convergence_test_deri_5p} and the covariance matrix $N_{\rm cov}$ in figure~\ref{fig:Convergence_test_Ncov}. For the convergence w.r.t. $N_{deri}$, we compare the forecasts calculated with the standard, compressed, and combined estimator described in Section~\ref{sec:fisher} for the five parameters $\Omega_m$, $h$, $\sigma_8$, $M_{\nu}$, and $|f_{R_0}|^{\lg 2}$ on both small and large scales. To calculate the constraints as a function of $N_{deri}$, we first randomly choose $N_{deri}$ indices from 0 to 499 without replacement. Then for the combined estimator only, the $N_{deri}$ indices are randomly divided in half: one half for the standard estimator and the other half for the compressed estimator. Finally, the derivatives of the statistics are estimated using the simulations at these indices. Therefore, the fluctuations seen between the constraint at $N_{deri,1}$ and $N_{deri,2}$ can stem from the randomness of drawing samples (and the random process in the further division of samples for the combined estimator).

For $P_{0,2,4}$, the results from the three estimators agree with each other very well on both small and large scales when $N_{deri}>100$. On the other hand, for the same estimator, the forecasts calculated with different $N_{deri}$ agree with each other as well. Therefore, figure~\ref{fig:Convergence_test_deri_5p} provides a dual convergence check for the Fisher matrix analysis. For $W_0+W_1^{0,2}+W_2^{0,2}+W_3$, differences can be seen even for $N_{deri}=500$ in the constraints on the four parameters except $|f_{R_0}|^{\lg 2}$ from the three estimators, but the constraints from the standard and compressed estimator provide an approximate lower and upper bound for the forecasts \cite{2023arXiv230508994C}. In addition, the constraints from the combined estimator for different $N_{deri}$ agree with each other when $N_{deri}>100$, even though there exist some fluctuations for $h$, $\sigma_8$ and $M_{\nu}$ in the constraints on large scales. For the combination of the power spectrum and MTs, the convergence is dominated by that of the MTs, we thus see very similar convergence curves to those of the MTs.

A great convergence with respect to $N_{cov}$ can be seen in figure~\ref{fig:Convergence_test_Ncov}, the ratio $\sigma_{\theta}(N_{cov})/\sigma_{\theta}(N_{cov}=5000)$ is smaller than $7.5\%$ ($2.5\%$) for both statistics and their combination on both small and large scales for all parameters when $N_{cov}>1000$ ($N_{cov}>3000$). 

As shown in figure~\ref{fig:Fisher_correlation}, both the degeneracy among $\Omega_b$, $h$ and $n_s$ and between $\sigma_8$ and $M_{\nu}$ are a potential source of instability when inverting the Fisher matrix to get parameter constraints. And we have shown in figure~\ref{fig:Convergence_test_deri_5p} if we rest the degeneracy among $\Omega_b$, $h$ and $n_s$ by fixing $\Omega_b$ and $n_s$ and only including $h$ in the Fisher analysis, we can already get converged forecasts for all cases considered in this work. In figure~\ref{fig:Convergence_deri_4p}, we further rest the degeneracy between $\sigma_8$ and $M_{\nu}$ and only perform the Fisher analysis for either $\{\Omega_m,\ h,\ M_{\nu},\ |f_{R_0}|^{\lg 2}\}$ or $\{\Omega_m,\ h,\ \sigma_8,\ |f_{R_0}|^{\lg 2}\}$ with the MTs. We find the results from the three estimators agree with each other well when $N_{deri}>300$ for the two Fisher analyses performed. For the left panel, the derivative w.r.t. $M_{\nu}$ is estimated with the highest order of differentiation scheme, and converged results can still be obtained even though the highest order estimator of the $M_{\nu}$ derivative is most prone to noise.  

\begin{figure}[tbp]
	\centering
	\includegraphics[width=1.0\textwidth]{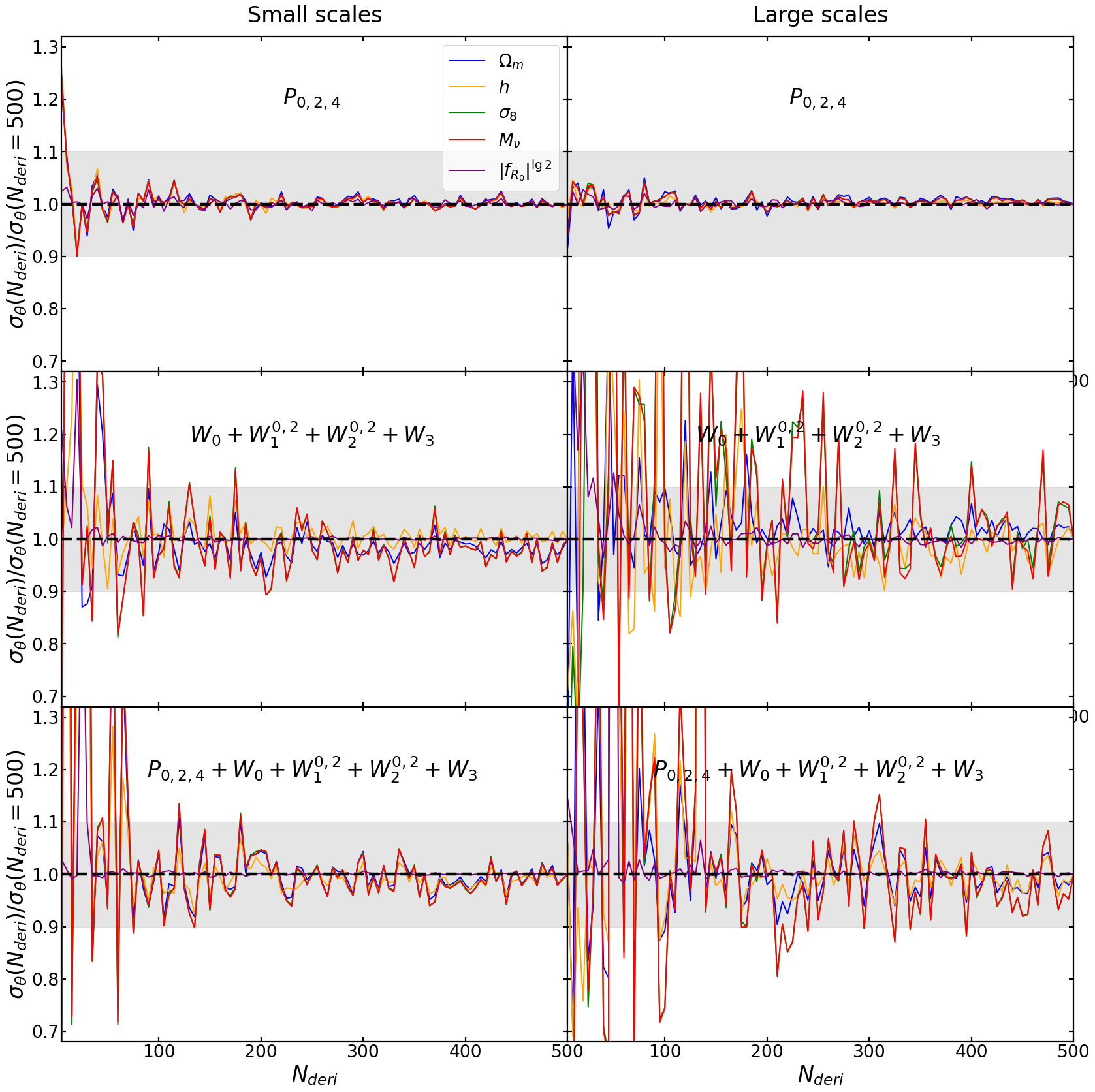}
	\caption{\label{fig:conver_ratio} The ratio of the marginalized errors obtained with derivatives estimated from $N_{deri}$ simulations to those from $N_{deri}=500$ simulations. The convergence is plotted for the power spectrum multipoles ($P_{0,2,4}$, first row), the MTs ($W_0+W^{0,2}_{1}+W^{0,2}_{1}+W_3$, second row), and their combination ($P_{0,2,4}+W_0+W^{0,2}_{1}+W^{0,2}_{1}+W_3$, third row) on $\Omega_{m}$, $h$, $\sigma_{8}$, $M_{\nu}$ and $f_{R_0}$ (blue, orange, green, red, purple) for both small ($k_{\rm{max}}=0.5~h\rm{Mpc}^{-1},\ R_G=5~h^{-1}\rm{Mpc}$, left column) and large ($k_{\rm{max}}=0.25~h\rm{Mpc}^{-1},\ R_G=10~h^{-1}\rm{Mpc}$, right column) scales.  The calculations are all based on covariance matrices estimated with $N_{cov}=5000$ simulations. Horizontal dashed lines correspond to $\sigma_{\theta}(N_{deri})/\sigma_{\theta}(N_{deri}=500)=1$ and grey shaded areas represent the $10\%$ fluctuations around $\sigma_{\theta}(N_{deri})/\sigma_{\theta}(N_{deri}=500)=1$.} 
\end{figure}

We would like to emphasize once again that we follow the method proposed in \cite{2023arXiv230508994C} to combine the compressed estimator with the standard estimator of the Fisher matrix, as this combined estimator is intended to provide less biased (if not unbiased) forecasts. By doing so, we aim to avoid overestimating the constraining power of the statistics, a well-known issue with the standard estimator when the noise in the derivative is large \cite{2021JCAP...04..029H,2022arXiv221012743H}.

However, as shown in Figure~\ref{fig:conver_ratio}, the combined estimator exhibits its own convergence issue: the ratio $\sigma_{\theta}(N_{deri})/\sigma_{\theta}(N_{deri}=500)$ fluctuates around 1 as the number of simulations used to estimate the derivatives, $N_{deri}$, varies. In general, we find that the ratios for all parameters, for both statistics and their combination, fluctuate around one, showing almost no bias. The way the amplitude of these fluctuations decreases with increasing 
$N_{deri}$ can be seen as a convergence test for the combined estimator of the Fisher matrix \footnote{Reliable error bars can only be added for the ratios at $N_{deri}<50$, where at least ten independent calculations of the ratio can be performed. For $N_{deri}>200$, which is more relevant to our analysis, only one or two independent calculations are available, so we cannot estimate error bars in this range.}. On small scales, we observe very small variations in the ratios for the power spectrum, with amplitudes less than $5\%$ when $N_{deri}\gtrsim 100$. For the MTs, the uncertainties are generally larger, but fluctuations remain under $10\%$ for $N_{deri}\gtrsim 200$. The convergence for the combination of the power spectrum and the MTs is similar to that of the MTs;  with uncertainties also reduced to below $<10\%$ for $N_{deri}\gtrsim 200$. 

On large scales, the fluctuations for the power spectrum are also small for all five parameters. However, for the MTs, the variations in the ratio for $M_{\nu}$ and $\sigma_8$ can still exceed $10\%$, even when $N_{deri}$ is close to 500. For the other three parameters, the fluctuations are below $10\%$ when $N_{deri}\gtrsim 200$. We observe similar but somewhat mitigated fluctuations for $P_{0,2,4}+W_0+W^{0,2}_{1}+W^{0,2}_{1}+W_3$.

In conclusion, although we have made efforts to avoid overestimating the constraining power of the statistics using the combined estimator for the Fisher matrix, we find that this estimator faces a different convergence challenge: large fluctuations can occur in the constraints, particularly for $M_{\nu}$ and $\sigma_8$ from the MTs on large scales. A large number of simulations would be required for these constraints to converge below the $5\%$ level. However, we find the fluctuations in the constraint on $|f_{R_0}|^{\lg 2}$ remain comfortably below $\sim 2\%$ for all the scales and statistics shown when $N_{deri}\gtrsim 200$.  This suggests that our main result and conclusion regarding the f(R) parameter are not likely to be significantly affected by the convergence issue of the Fisher matrix.

\section{Convergence w.r.t. inherent error of numerical differentiation methods}
\label{sec:test_deri}
\begin{figure}[tbp]
	\centering
	\includegraphics[width=1.0\textwidth]{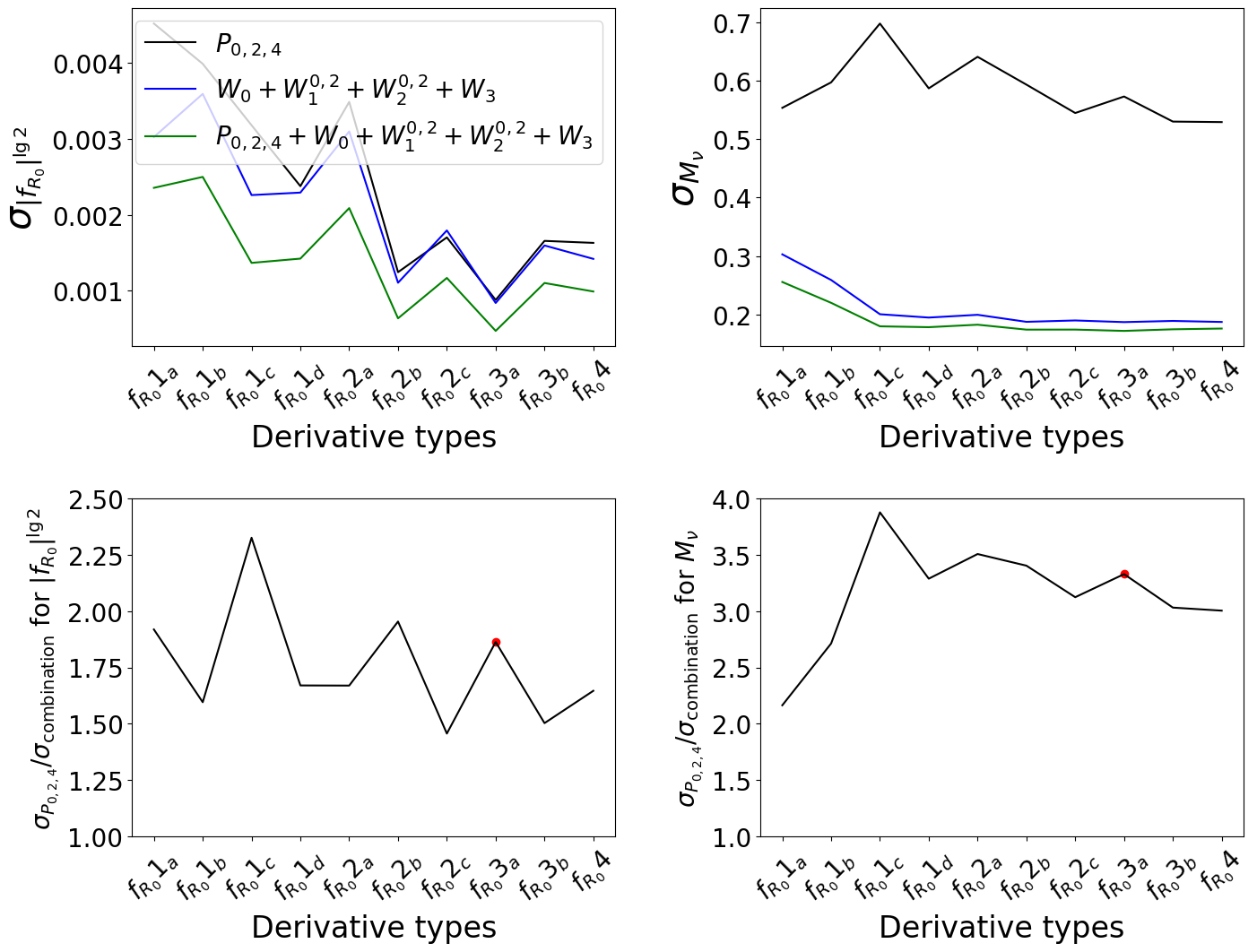}
	\caption{\label{fig:deri_conver} Top panels: convergence of the marginalized error as a function of the estimator type of the $f_{R_0}$ derivative. Results are shown for the power spectrum (black), MTs (blue), and combination of $P_{0,2,4}$ and MTs (green) for $|f_{R_0}|^{\lg 2}$ (top left panel) and $M_{\nu}$ (top right panel). Bottom panels: ratio of the marginalized error for the power spectrum to that for the combination of statistics for $|f_{R_0}|^{\lg 2}$ (bottom left panel) and $M_{\nu}$ (bottom right panel). Red points denote the ratio that we report in our main results.} 
\end{figure}

Two different kinds of errors exist in the estimate of derivatives with cosmological simulations: the first one comes from the cosmic variance, which is the fluctuations of observables among different realizations of the simulations; the second one is the inherent error of the numerical differentiation methods, it depends on the step size and the order of the differentiation method. We have discussed the first kind of error in the Appendix\ref{sec:conver} and will investigate the second kind here.

Although we note that the $f_R^{+}$ model is so close to fiducial cosmology that the difference in statistics between $f_R^{+}$ and GR is smaller than the cosmic variance for most of the bins, we will still exploit it here for a comprehensive test of the convergence w.r.t. the inherent error of numerical differentiation methods. With the simulations list in Table~\ref{tab:s}, we can estimate the derivatives w.r.t. $f_{R_0}$ with:
\begin{equation}
	f_{R_0}1_a:\frac{\partial \boldsymbol{\mu}}{\partial f_{R_0}}=\frac{\boldsymbol{\mu}(8df_{R_0})-\boldsymbol{\mu}(\theta^{ZA}_{fid})}{8df_{R_0}},
\end{equation}
\begin{equation}
	f_{R_0}1_b:\frac{\partial \boldsymbol{\mu}}{\partial f_{R_0}}=\frac{\boldsymbol{\mu}(4df_{R_0})-\boldsymbol{\mu}(\theta^{ZA}_{fid})}{4df_{R_0}},
\end{equation}
\begin{equation}
	f_{R_0}1_c:\frac{\partial \boldsymbol{\mu}}{\partial f_{R_0}}=\frac{\boldsymbol{\mu}(2df_{R_0})-\boldsymbol{\mu}(\theta^{ZA}_{fid})}{2df_{R_0}},
\end{equation}
\begin{equation}
	f_{R_0}1_d:\frac{\partial \boldsymbol{\mu}}{\partial f_{R_0}}=\frac{\boldsymbol{\mu}(df_{R_0})-\boldsymbol{\mu}(\theta^{ZA}_{fid})}{df_{R_0}},
\end{equation}
\begin{equation}
	f_{R_0}2_a:\frac{\partial \boldsymbol{\mu}}{\partial f_{R_0}}=\frac{-\boldsymbol{\mu}(8df_{R_0})+4\boldsymbol{\mu}(4df_{R_0})-3\boldsymbol{\mu}(\theta^{ZA}_{fid})}{8df_{R_0}},
\end{equation}
\begin{equation}
	f_{R_0}2_b:\frac{\partial \boldsymbol{\mu}}{\partial f_{R_0}}=\frac{-\boldsymbol{\mu}(4df_{R_0})+4\boldsymbol{\mu}(2df_{R_0})-3\boldsymbol{\mu}(\theta^{ZA}_{fid})}{4df_{R_0}},
\end{equation}
\begin{equation}
	f_{R_0}2_c:\frac{\partial \boldsymbol{\mu}}{\partial f_{R_0}}=\frac{-\boldsymbol{\mu}(2df_{R_0})+4\boldsymbol{\mu}(df_{R_0})-3\boldsymbol{\mu}(\theta^{ZA}_{fid})}{2df_{R_0}},
\end{equation}
\begin{equation}
	f_{R_0}3_a:\frac{\partial \boldsymbol{\mu}}{\partial f_{R_0}}=\frac{\boldsymbol{\mu}(8df_{R_0})-12\boldsymbol{\mu}(4df_{R_0})+32\boldsymbol{\mu}(2df_{R_0})- 21\boldsymbol{\mu}(\theta^{ZA}_{fid})}{12df_{R_0}},
\end{equation}
\begin{equation}
	f_{R_0}3_b:\frac{\partial \boldsymbol{\mu}}{\partial f_{R_0}}=\frac{\boldsymbol{\mu}(4df_{R_0})-12\boldsymbol{\mu}(2df_{R_0})+32\boldsymbol{\mu}(df_{R_0})- 21\boldsymbol{\mu}(\theta^{ZA}_{fid})}{12df_{R_0}},
\end{equation}
\begin{equation}
	f_{R_0}4:\frac{\partial \boldsymbol{\mu}}{\partial f_{R_0}}=\frac{-\boldsymbol{\mu}(8df_{R_0})+28\boldsymbol{\mu}(4df_{R_0})-224\boldsymbol{\mu}(2df_{R_0})+512\boldsymbol{\mu}(df_{R_0})- 315\boldsymbol{\mu}(\theta^{ZA}_{fid})}{168df_{R_0}},
\end{equation}
where $df_{R_0}=0.0127$, and $\boldsymbol{\mu}(8df_{R_0})$, $\boldsymbol{\mu}(4df_{R_0})$, $\boldsymbol{\mu}(2df_{R_0})$, $\boldsymbol{\mu}(df_{R_0})$, and $\boldsymbol{\mu}(\theta^{ZA}_{fid})$ are estimated with 500 realizations for model $f_R^{++++}$, $f_R^{+++}$, $f_R^{++}$,  $f_R^{+}$, and Fiducial ZA, respectively.

In Figure~\ref{fig:deri_conver}, we examine how the marginalized errors on $|f_{R_0}|^{\lg 2}$ and $M_{\nu}$ change when the $f_{R_0}$ derivatives are calculated using different estimators. From left to right, the inherent error of these estimators progressively decreases. For $|f_{R_0}|^{\lg 2}$, we find that the constraints from $P(k)$, the MTs, and their combination all begin to converge at $f_{R_0}2_b$. The convergence of constraints on other parameters follows a similar trend to that of $M_{\nu}$, so we focus on the results for $M_{\nu}$ here. The constraints from the power spectrum start to converge at $f_{R_0}2_c$, while those from the MTs and the combined statistics converge earlier, at $f_{R_0}1_c$. We also highlight the improvements in the constraints from the combination of statistics on both $|f_{R_0}|^{\lg 2}$ and $M_{\nu}$. For both parameters, the variations in the improvements diminish after $f_{R_0}1_d$.

It is worth noting that the estimator ($f_{R_0}3_a$) used for our main results may slightly overestimate the improvement in the constraining power of the power spectrum. We have opted for this estimator because it shows better convergence with respect to noise in the derivatives compared to estimator $f_{R_0}4$. Nonetheless, the difference in both the Fisher forecasts and their convergence between estimator $f_{R_0}3_a$ and $f_{R_0}4$ is not significant.

\section{A more conservative result}
\label{sec:conservative_result}

\begin{figure}[tbp]
	\centering 
	\includegraphics[width=1.0\textwidth]{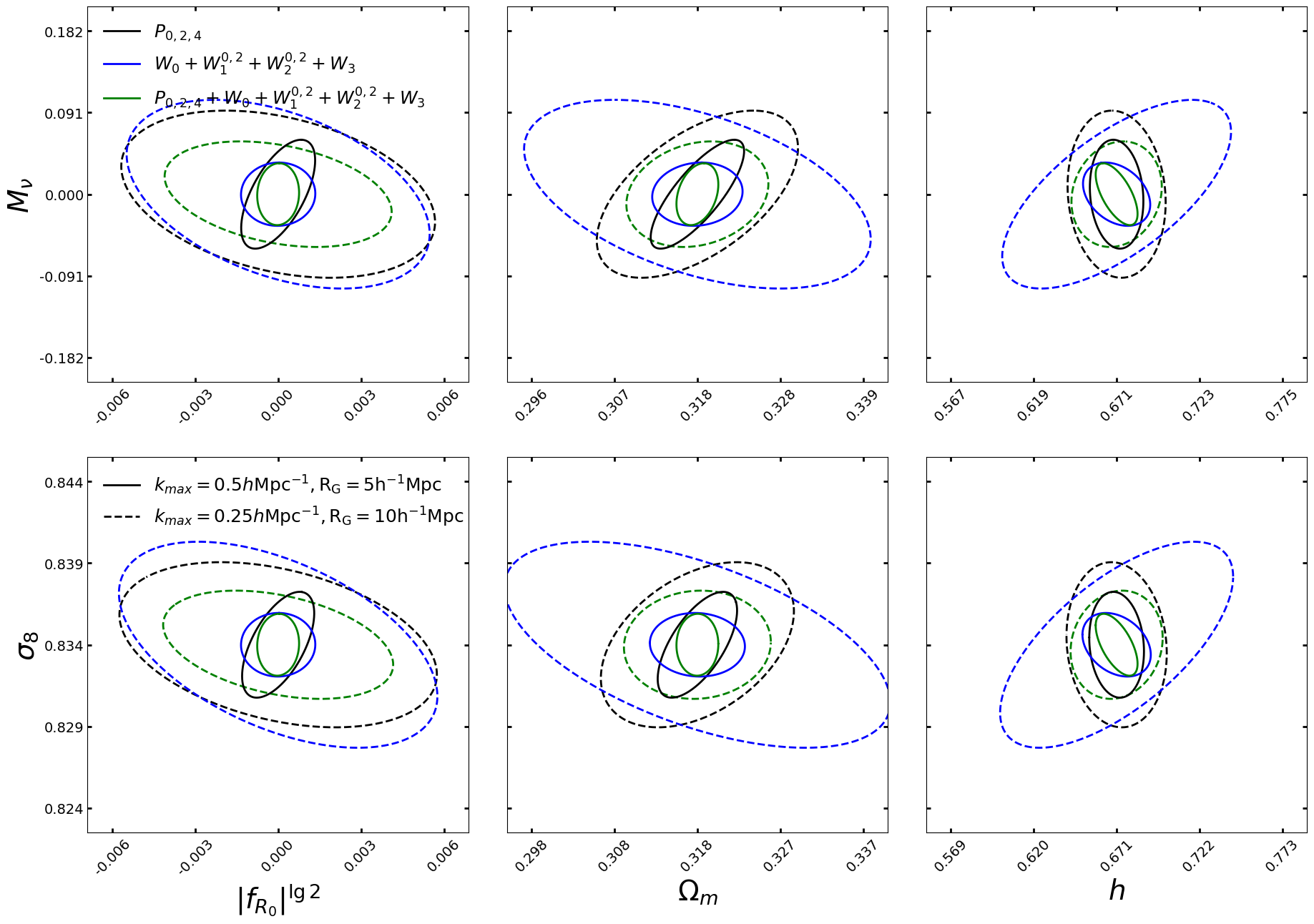}
	\caption{\label{fig:Conservative_results} Parameter constraint contours for $\{\Omega_m,\ h,\ M_{\nu},\ |f_{R_0}|^{\lg 2}\}$ (first row) and $\{\Omega_m,\ h,\ \sigma_8,\ |f_{R_0}|^{\lg 2}\}$ (second row). 
Results for the power spectrum $P_{0,2,4}$, the MTs $W_0+W^{0,2}_1+W^{0,2}_2+W_3$, and the combination of power spectrum and MTs $P_{0,2,4}+W_0+W^{0,2}_1+W^{0,2}_2+W_3$ are plotted with black, blue, and green lines, respectively.  Solid and dashed lines for small scales ($k_{\rm{max}}=0.5~h\rm{Mpc}^{-1},R_G=5~h^{-1}\rm{Mpc}$) and large scales ($k_{\rm{max}}=0.25~h\rm{Mpc}^{-1},R_G=10~h^{-1}\rm{Mpc}$).}
\end{figure}

The strong degeneracy between $\sigma_8$ and $M_{\nu}$ can make the Fisher matrix ill-conditioned and the noise in the derivatives and covariance matrix may be significantly amplified. Although we have done the convergence tests in Appendix~\ref{sec:conver} to show converged results can be obtained even when the $\sigma_8$ and $M_{\nu}$ degeneracy is included in the Fisher matrix analysis, we will still present a more conservative Fisher forecast in this appendix.

In figure~\ref{fig:Conservative_results}, we provide the parameter confidence contours when the Fisher matrix analysis is done for either $\{\Omega_m,\ h,\ M_{\nu},\ |f_{R_0}|^{\lg 2}\}$ or $\{\Omega_m,\ h,\ \sigma_8,\ |f_{R_0}|^{\lg 2}\}$. In the first case, constraints from the MTs are weaker on $\Omega_m$, $h$, and $M_{\nu}$, but stronger on $|f_{R_0}|^{\lg 2}$ than those from $P_{0,2,4}$ on large scales.  The constraining power of the MTs is only slightly weaker on $h$ than $P_{0,2,4}$ on small scales. In the second case, smaller errors can be seen from $P_{0,2,4}$ on all the four parameters on large scales; but the MTs are more sensitive to $\sigma_8$ and $|f_{R_0}|^{\lg 2}$ on small scales. No matter in which case, on which scale, complementary information can be found from the MTs to $P_{0,2,4}$. Degeneracies are broken and tighter constraints are obtained with the combination of the power spectrum and MTs.

Comparing figure~\ref{fig:Conservative_results} with figure~\ref{fig:individual}, we find the constraining power of $P_{0,2,4}$ is mainly limited by the $\sigma_8$ and $M_{\nu}$ degeneracy, and much tighter constraints could be obtained if the degeneracy is broken. The small-scale MTs are very sensitive to $M_{\nu}$ and $\sigma_8$, and tight constraints can be obtained no matter the degeneracy between $M_{\nu}$ and $\sigma_8$ is on or off. They are also sensitive to $|f_{R_0}|^{\lg 2}$, so tighter or comparable constraints can be derived in general from the MTs compared with those from $P_{0,2,4}$.

\section{The influence of the explicit ``k-cut'' on the constraints}
\label{sec:k-cut}

\begin{center}
	\small
	\begin{table}[tbp]
		\scalebox{1}[1]{
                    \begin{tabular}{|c|c|c|c|c|}
                    \hline
                    Statistic & MFs surface & MFs grid & MFs surface & MFs grid \\
                    \hline
                    Scale & \multicolumn{2}{|c|}{$ R_G=5~h^{-1}\rm{Mpc}$} & \multicolumn{2}{|c|}{$ R_G=10h^{-1}\rm{Mpc}$} \\
                    \hline
                    $\Omega_m$ & 0.006 & 0.006  & 0.022 & 0.023 \\
                    $h$ & 0.019 & 0.019 & 0.070 & 0.073 \\
                    $\sigma_8$ & 0.011 & 0.011 & 0.016 & 0.017 \\
                    $M_\nu$ & 0.189 & 0.188  & 0.285 & 0.286 \\
                    $|f_{R_0}|^{\lg 2}$ & 0.0010 & 0.0010 & 0.0044 & 0.0044 \\

                    \hline
                    \end{tabular}
		}
		\caption{\label{tab:MFs_surface_vs_grid} Comparison between the marginalized parameter constraints from the MFs measured with two different algorithms for two smoothing scales (the first two columns for $ R_G=5~h^{-1}\rm{Mpc}$ while the last two columns for $R_G=10~h^{-1}\rm{Mpc}$). The computationally expensive method based on the bounding surfaces generated for excursion sets is denoted as ``MFs surface'' and the efficient grid-based Crofton’s formula is denoted as ``MFs grid''.} 
	\end{table}
\end{center}

\begin{center}
	\small
	\begin{table}[tbp]
		\scalebox{1}[1]{
                    \begin{tabular}{|c|c|c|c|c|}
                    \hline
                    Statistic & MFs & MFs k-cut & MFs & MFs k-cut \\
                    \hline
                    Scale & \multicolumn{2}{|c|}{$ R_G=5~h^{-1}\rm{Mpc}$} & \multicolumn{2}{|c|}{$ R_G=10h^{-1}\rm{Mpc}$} \\
                    \hline
                    $\Omega_m$ & 0.006 & 0.006  & 0.023 & 0.027 \\
                    $h$ & 0.019 & 0.020 & 0.073 & 0.082 \\
                    $\sigma_8$ & 0.011 & 0.011 & 0.017 & 0.016 \\
                    $M_\nu$ & 0.188 & 0.184  & 0.287 & 0.271 \\
                    $|f_{R_0}|^{\lg 2}$ & 0.0010 & 0.0011 & 0.0044 & 0.0045 \\

                    \hline
                    \end{tabular}
		}
		\caption{\label{tab:MFs_allk_vs_cutk} Comparison between the marginalized parameter constraints from the MFs with or without the ``k-cut'' (the first two columns for $R_G=5~h^{-1}\rm{Mpc}$ while the last two columns for $R_G=10~h^{-1}\rm{Mpc}$). For $R_G=5~h^{-1}\rm{Mpc}$ and $R_G=10~h^{-1}\rm{Mpc}$, the modes $k>0.5 h\rm Mpc^{-1}$ and $k>0.25 h\rm Mpc^{-1}$ are removed before the Gaussian smoothing procedure, respectively.} 
	\end{table}
\end{center}

The MTs are measured through time-consuming computations over the complicated triangulated surfaces of density thresholds, making the process extremely resource-intensive.  For example, it can take 8 to 12 CPU hours to process one $512\times 512\times 512$ density field with $R_G=5h^{-1} \rm Mpc$, depending on the cosmological model used in the N-body simulations. Performing another Fisher analysis with the "k-cut" applied would require over $10^5$ CPU hours. In contrast, the grid-based measurement of the MFs using Crofton's formula, as proposed by \cite{1997ApJ...482L...1S}, is significantly faster. It takes less than 50 CPU seconds for a single $512\times 512\times 512$ density field, making the grid-based MFs measurement more than 700 times faster than the surface-based MTs. Additionally, as shown in table~\ref{tab:MFs_surface_vs_grid}, the constraints obtained from the grid-based MFs are comparable to those from the surface-based MFs. Therefore, we have chosen to compare the results with and without the explicit "k-cut" using these fast grid-based MF measurements.

In Table~\ref{tab:MFs_allk_vs_cutk}, we find the explicit ``k-cut'' only slightly weakens the constraints on $h$ and $|f_{R_0}|^{\lg 2}$ for $ R_G=5~h^{-1}\rm{Mpc}$. The constraints on other parameters are barely affected after removing the modes $k>0.5 h\rm Mpc^{-1}$. For $ R_G=10~h^{-1}\rm{Mpc}$, the constraining power of the MFs on $\Omega_m$ and $h$ is reduced by $17\%$ and $12\%$,  respectively,  after cutting out the modes $k>0.25 h\rm Mpc^{-1}$. However, the constraints on other parameters including, $|f_{R_0}|^{\lg 2}$, are only slightly impacted by the cut in $k$-modes. Overall, the results in Table~\ref{tab:MFs_allk_vs_cutk} are consistent with our expectation that the "k-cut" has minimal impact on our Fisher forecasts.

\bibliographystyle{JHEP}
\bibliography{MTMGNU}
\end{document}